\documentclass[conference]{IEEEtran}

\usepackage{cite}
\usepackage{amsmath}
\usepackage{xspace}
\usepackage[export]{adjustbox}
\usepackage{wrapfig}
\usepackage{subfig}
\usepackage{float}
\usepackage{algorithm}
\usepackage[noend]{algpseudocode}
\usepackage{comment}
\usepackage{xcolor}
\usepackage{url}
\usepackage{fancyhdr}
\usepackage[font=small,labelfont=bf]{caption}

\def\BibTeX{{\rm B\kern-.05em{\sc i\kern-.025em b}\kern-.08em
    T\kern-.1667em\lower.7ex\hbox{E}\kern-.125emX}}

\DeclareMathOperator*{\argmin}{arg\,min}

\newcommand{\eat}[1]{}
\newcommand{\para}[1]{\smallskip \noindent {\bf #1}}
\newcommand{\softpara}[1]{\smallskip \noindent \underline{#1}}

\newcommand{\A}{\ensuremath{\mathcal{A}}\xspace}

\newcommand{\PUParam}{\ensuremath{\mathcal{R}}\xspace}

\newcommand{\Sensor}{\ensuremath{\mathcal{S}}\xspace}
\newcommand{\sii}{\ensuremath{\mathrm{S_i}}\xspace}
\newcommand{\rj}{\ensuremath{R_{j}}\xspace}

\newcommand{\loc}{\ensuremath{l}\xspace}

\newcommand{\pset}{{\tt PU-Setting}\xspace}
\newcommand{\sset}{{\tt SS-Setting}\xspace}
\newcommand{\nn}{{\tt NN}\xspace}
\newcommand{\svr}{{\tt SVR}\xspace}

\newcommand{\ip}{{\tt IP-Based}\xspace}
\newcommand{\perr}{\ensuremath{\mathrm{\mathcal{A}_{err}}}\xspace}
\newcommand{\pfp}{\ensuremath{\mathrm{\mathcal{A}_{fp}}}\xspace}
\newcommand{\afp}{\ensuremath{\alpha_{FP}}\xspace}
\newcommand{\afn}{\ensuremath{\alpha_{FN}}\xspace}

\newcommand{\plname}{\ensuremath{\rho}}                        
\newcommand{\plfunc}[2]{\plname\left(#1,#2\right)}

\newcommand{\blue}[1]{\textcolor{black}{#1}}
\newcommand{\bluee}[1]{\textcolor{black}{#1}}
\newcommand{\bleu}[1]{\textcolor{black}{#1}}

\newcommand{\iab}[1]{\textcolor{black}{#1}}
\newcommand{\iar}[1]{\textcolor{black}{#1}}
\newcommand{\iam}[1]{\textcolor{black}{#1}}

\newcommand{\name}{{\tt DeepAlloc}\xspace}
\newcommand{\sname}{{\tt SH-Alloc}\xspace}
\newcommand{\nameg}{{\tt DeepAlloc-Greedy}\xspace}

\newcommand{\namenn}{{\tt DeepAlloc-NN}\xspace}
\newcommand{\namernn}{{\tt DeepAlloc-RNN}\xspace}
\newcommand{\bina}{{\tt Binary-Alloc}\xspace}

\newcommand{\rdtc}[1]{\textcolor{black}{#1}}
\newcommand{\mgtc}[1]{\textcolor{black}{#1}}

\newcommand{\tmagenta}{\color{black}}
\newcommand{\tblack}{\color{black}}
\newcommand{\sj}{\ensuremath{s_{j}}\xspace}

\begin{document}

\title{\name: Deep Learning Approach to Spectrum Allocation in Shared Spectrum Systems}

\author{\IEEEauthorblockN{Mohammad Ghaderibaneh, Caitao Zhan, Himanshu Gupta}
\IEEEauthorblockA{\textit{Department of Computer Science} \\
\textit{Stony Brook University, USA}\\
}}

\maketitle

\begin{abstract}
Shared spectrum systems facilitate spectrum allocation to unlicensed users without harming the licensed users; they offer great promise in optimizing spectrum utility, 
but their management (in particular, efficient spectrum allocation to unlicensed users)
is challenging.
To allocate spectrum efficiently to secondary users (SUs) in general scenarios, we 
fundamentally need to have knowledge of the signal path-loss function. In practice, however, even the best-known path-loss models have unsatisfactory accuracy, and conducting extensive surveys to gather path-loss values is infeasible.
\mgtc{Thus, the current allocation methods are either (i) too conservative in preventing 
interference that they sacrifice performance, or (ii) are based on imperfect  
propagation models and/or spectrum sensing with insufficient spatial granularity.}
This leads to poor spectrum utilization, the fundamental objective of shared
spectrum systems.

\mgtc{In this work}, we thus propose to {\em learn} the spectrum allocation function directly using supervised learning techniques. \mgtc{Such an approach has the potential to deliver near-optimal performance with sufficient and effective training data. In addition, it has the advantage of being viable even when certain information is unavailable; e.g., in settings 
where PUs' information is not available, we make use of a crowdsourced 
sensing architecture and use the spectrum sensor readings as features.}
\mgtc{In general,} for spectrum allocation to a single SU, we develop a CNN-based approach (called \name) 
and address various challenges that arise in our context; to handle multiple SU
requests simultaneously, we extend our approach based on recurrent neural 
networks (RNNs).
Via extensive large-scale simulation and a small testbed, we demonstrate the effectiveness of our developed techniques; in particular, we observe that our approach improves the accuracy of
standard learning techniques and prior work by up to 60\%. 
\end{abstract}

\begin{IEEEkeywords}
Spectrum Sharing, Spectrum Allocation, Deep Learning, Convolutional Neural Networks
\end{IEEEkeywords}

\section{Introduction}

The RF spectrum is a natural resource in great demand due to the
unabated increase in mobile (and hence, wireless) data
consumption~\cite{Jeffrey14}. The research community has
addressed this capacity crunch via the development of {\em shared spectrum
  paradigms}, wherein the spectrum is made available to unlicensed
(Secondary) users as long as they do not interfere with the
transmission of licensed incumbents, i.e.,  primary users (PUs).  Effective management
(and in particular, allocation) 
of spectrum in such shared spectrum systems is challenging, and
several spectrum management architectures have been proposed over the
years~\cite{spectrumAllocationSurvey13,parishad18,milind05,sudeep16}.
A significant shortcoming of these architectures and methods is that
spectrum allocation is done very conservatively to ensure correctness,
or is based on imperfect propagation
modeling~\cite{chamberlin82,ayon14} or
spectrum sensing with poor spatial granularity. This leads to poor
spectrum utilization, the fundamental objective of the shared spectrum
systems. In this paper, we develop a learning-based approach to efficient
allocation of spectrum in such shared spectrum systems.

\para{Motivation and Overall Approach.} 
In general, to allocate spectrum efficiently to secondary users, 
we fundamentally need to have knowledge of the signal path loss between 
largely arbitrary pair of points. In practice, however, even the best
known path-loss models~\cite{chamberlin82,hata2000} have
unsatisfactory accuracy, and conducting extensive surveys to gather
path-loss values are infeasible and, moreover, may not even reflect
real-time channel conditions.
To circumvent the above challenge, we propose to instead learn the
spectrum allocation function directly using supervised learning
techniques. 
In scenarios where primary-user (PU) parameters may not
be available (e.g., navy radars in CBRS band~\cite{fcc35band}),
we use a crowdsourced
sensing architecture where we utilize relatively low-cost spectrum
sensors independently deployed with a high
granularity~\cite{ayon17,calvo2017crowdsourcing,shibo2014crowdssensing,shi2014crowdsourcing}. A sensing architecture also 
enables spectrum allocation based on real-time channel
conditions~\cite{curran2019procsa}.

We propose to use supervised learning techniques to learn the 
Spectrum Allocation (SA) 
function with
the input (features) being the primary-user parameters, spectrum sensor (SS) readings, and secondary user (SU) request parameters, and the output (label) being the maximum power that can be allocated to the SU without
resulting in any harmful interference to the PUs' receivers. 
\mgtc{Based on the insight that the input to the SA function can be represented as an image and thus the SA function can be framed as an image regression 
problem, we develop a convolution neural network (CNN) model to learn
the allocation function as CNNs have been most successful 
learning models for image regression and classification tasks.
To develop an effective CNN-based learning architecture, we 
develop techniques to represent the inputs to the SA function
as an image, design an efficient CNN architecture, 
and address associated challenges.}

\para{Our Contributions.} We make the following contributions.
\begin{enumerate}
    \item 
    We motivate and 
    propose \iab{using} Convolution Neural Networks (CNNs) for efficient learning of the spectrum allocation function.
    %
    In particular, we develop an efficient CNN architecture based on {\em pre-training} a deep model using \iab{many} {\em auto-generated} samples/images, followed by training using samples gathered over the given region. 
    
    \item 
    \iam{We develop a novel technique to represent the spectrum allocation function input (i.e., the location and transmission/received powers of primary users or spectrum sensors, and the request parameters of the secondary user) as an image; such an image representation is essential to effectively use a CNN-based learning model. In addition, we develop techniques to minimize false positive errors, handle multi-path effects, and further improve accuracy via synthetic samples, in our context of a CNN-based learning approach.}
    
    \item 
    To allocate spectrum simultaneously to multiple concurrent SUs, we develop a deep-learning architecture based on recurrent neural networks (RNNs) with inputs from our CNN-based architecture for single SUs.
    
    \item 
    We evaluate our techniques using large-scale simulations \iab{and} an outdoor testbed and demonstrate the effectiveness of our developed techniques. We observe that our approach improves the accuracy of other approaches by up to 60\%.
\end{enumerate}

\para{Paper Organization.} The rest of the paper is organized as follows. In the following section, we develop our spectrum allocation model and setting, discuss related work, and give a high-level overview of our approach. In \S\ref{sec:deep_learning}, we develop our CNN-based deep learning model and associated techniques for spectrum allocation. We discuss our simulation results in \S\ref{sec:simulations}, and end with concluding remarks in \S\ref{sec:conc}.

\section{\mgtc{Model,} Related Work, and Our Approach}
\label{sec:related}

\begin{figure}[h]
    \includegraphics[width=0.8\linewidth, center]{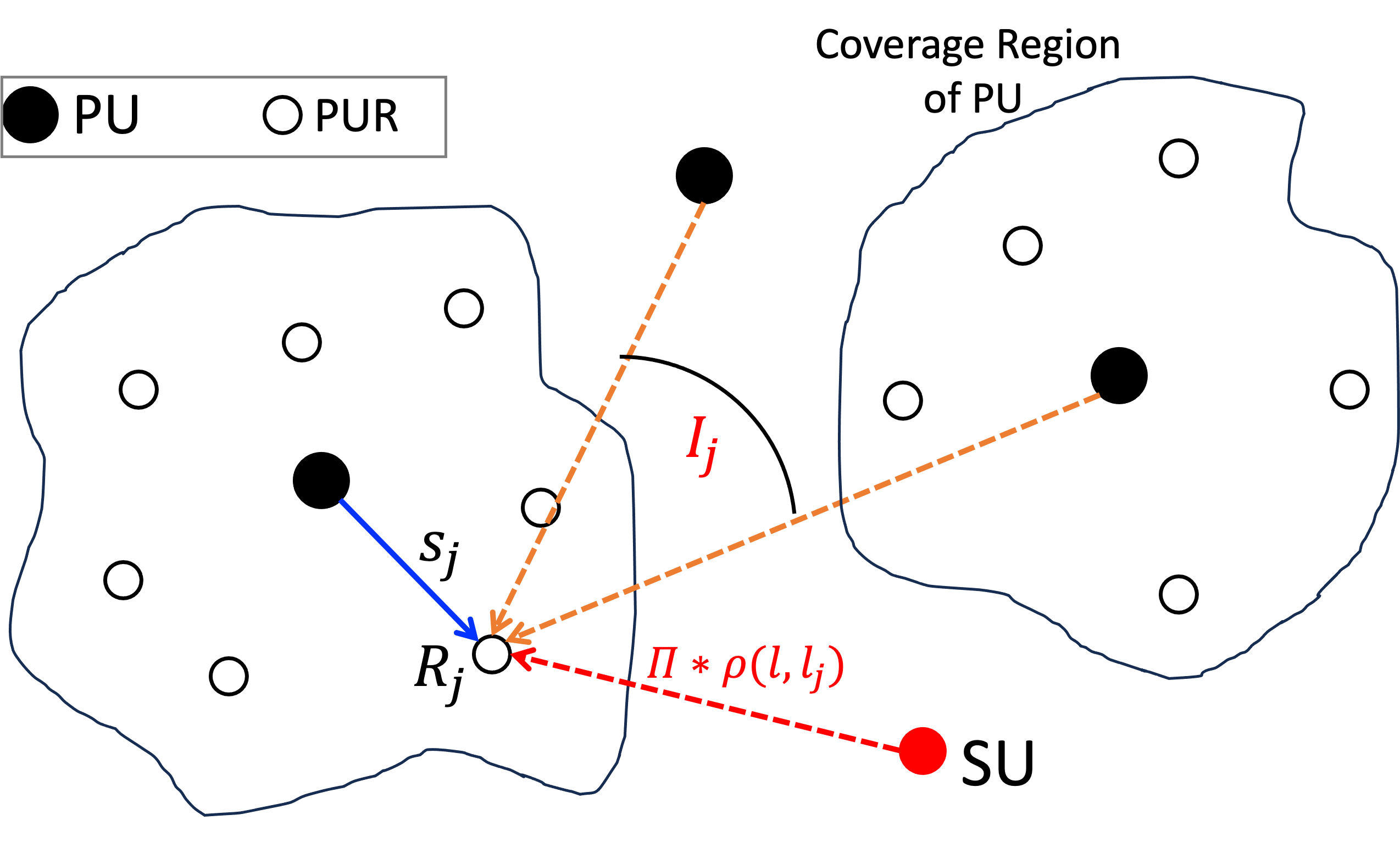} 
    \caption{\mgtc{Eqn.~(\ref{eq:maxpower}) Illustration. \iab{The} 
    optimal power that
    can be allocated to an SU is such that, at each PUR (a PU's receiver)
    the signal-to-noise ratio is more than \iab{the} desired ratio, $\beta$.
    Above, \rj is a certain PUR, $I_j$ is the total interference at \rj from other PUs,
    \sj is the signal strength received at \rj from its PU, and $\Pi \cdot \rho(\loc, l_j)$ is the interference due to the SU at \rj where $\loc$ and $l_j$ are their respective locations.}}
    \label{fig:desired}
\end{figure}
\tmagenta
\para{Shared Spectrum and Entities.}
A shared spectrum system mainly consists of licensed primary users (PUs) 
and unlicensed secondary users (SUs) who make spectrum allocation requests 
to the centralized \textit{spectrum manager}.
A secondary user (SU) requests authorization to transmit with certain
desired parameters (e.g., location, duration, frequency, transmission power).
For a given SU request, the spectrum manager determines whether the SU's request
can be granted based on whether its transmission under granted parameters 
would cause harmful interference to any of the intended receivers of a 
PUs signal, as discussed below. Additional entities in a shared spectrum 
system of relevance to our model are PU receivers and spectrum sensors, as
described below.

\softpara{PU Receivers (PURs).}
A  way of modeling intended receivers of a PU could be to define a coverage region around PU wherein we wish to guarantee reception (see Fig.~\ref{fig:desired}). 
In this work, as in~\cite{gupta19,anh10}, we model the intended 
receivers of a PU as a finite set of receiver nodes, which we denote as 
PURs. The PUR model is without loss of generality since, in general, the
PURs can be distributed arbitrarily around a PU. 
The PUR model is also more computationally efficient than the coverage-region
model in determining whether an SU's transmission causes harmful interference 
to PU's receivers. 
Later, we assume that the distribution method for PURs 
is ``similar'' across all PUs, to obviate the need to represent 
them explicitly in the SA's input image 
(see~\S\ref{sec:basic-cnn} for more details).


\para{Spectrum Allocation Objective: SU Transmission Power.}
The general spectrum allocation problem is to allocate optimal power to an SU's request across spatial, frequency, and temporal domains. We focus on the core function approximation problem, which is to determine the optimal power allocation to an SU for a \underline{given} location, channel, and time instant---since frequency and temporal domains are essentially ``orthogonal'' dimensions of the problem and thus can be easily handled independently (as done in~\S\ref{sec:multiple_sus}). We thus assume a single channel and instant for now, and discuss multiple channels and request duration in~\S\ref{sec:multiple_sus}.

\para{Determining Optimal Power Allocation.}
Consider a shared spectrum area with PUs deployed, and an SU request for transmission from a given location $\loc$. 
{\em If the PUs information and the path-loss function are known}, then 
the optimal power that can be allocated to SU without causing harmful interference to PUs can be computed as 
follows~\cite{curran2019procsa,dou17,gupta19}.
Let us denote the path loss function between a pair of given locations by 
$\plfunc{}{}$; thus, a signal transmitted at power $t_i$ from location $l_i$
yields a received power of $t_i \cdot \plfunc{l_i}{l_j}$ at location $l_j$.
Let the total interference from other PUs plus noise at \rj be $I_j$.
To ensure that the signal-to-noise ratio at each \rj is more than \iab{the desired}
value, say $\beta$, the maximum power $\Pi$ that can be allocated to the 
SU is:
\begin{equation}
\label{eq:maxpower}
   \Pi \leq \min_{j}\frac{(\sj/\beta) - I_j}{\rho(\loc, l_j)},
\end{equation}
where $\sj = t_i \rho(l_i, l_j)$ is the signal strength received at 
\rj from its PU transmitting at power $t_i$ from 
location $l_i$. See Fig.~\ref{fig:desired}. 
Note that the above formulation is largely without loss of generality---
as the path-loss function $\rho$ can be arbitrary. 
Irrespective, {\em the fundamental techniques developed in our work are largely independent of the formulation or algorithm used to determine the optimal allocation power}---since learning models and techniques are solely based on training examples. In \S\ref{sec:simulations}, we use the above formulation to generate the training examples for the models.


\para{Two Settings: \pset and \sset} 
The above formulation requires knowledge of PUs information as well as the 
path-loss functions. However, in most settings, neither of them may be available; in fact, this
is one of the key motivations of our learning approach. 
In particular, PU parameters may not be available in military or government
setting, e.g., in the CBRS 
3.5GHz shared band \cite{fcc35band} wherein the licensed users 
include Navy radar systems.
In light of \iab{the} above, we consider two different settings in this paper, 
based on the availability of PU information/parameters.

\softpara{\pset.} 
In this setting, the PUs' parameters are available---for determination of spectrum allocation. A PU's parameters include its location, transmit power, and its PURs' locations.
        
\softpara{\sset.}
In this setting, PUs' parameters may not be available, e.g., in military or 
government settings. In this case, to determine spectrum allocation, we 
make use of a crowdsourced sensing architecture where relatively low-cost spectrum
sensors (SS) are deployed with a high granularity~\cite{ayon17,calvo2017crowdsourcing,shi2014crowdsourcing,shibo2014crowdssensing}. 
In such a crowdsourced sensing architecture, allocation decision is based on SS parameters, which includes \iab{each sensor's} location and received (aggregated) 
signal strength from the PUs (\iab{presumably} representative of the PUs parameters).
Allocation based on SSs parameters is implicitly based on {\em real-time} channel conditions, which is important for accurate and optimized spectrum allocation as the conditions affecting signal attenuation (e.g., air, rain, vehicular traffic) may change over time.
\tblack

\subsection{Related Work}
\label{sub:related}

The spectrum allocation problem has been studied extensively 
(see~\cite{qinq07, spectrumAllocationSurvey13, manal17} for \iab{a survey}), 
especially in the context of shared spectrum systems. 
In a centralized SM
architecture, it is generally assumed that the SM has complete
knowledge of the PU parameters. Many prior works also assume a
propagation model which, in conjunction with known PU parameters,
allows spectrum allocation power to be computed via linear
programming~\cite{spectrumAllocationSurvey13} or other simple techniques for common
optimization objectives.
However, in practice, PU parameters may not be available, e.g., in the
CBRS (3550-3700 MHs, in the 3.5GHz) band~\cite{fcc35band} wherein the
licensed users include Navy radar systems. As most propagation models have
unsatisfactory accuracy, spectrum allocation must be done overly
conservatively for correctness.
In particular, in TV white spaces spectrum (54-698 MHz)~\cite{fccwhite}, 
spectrum allocation is done based on a
database with TV channel availability at each location; in essence,
the SU is allowed to transmit with a certain power if the signal
received at its location is below a low threshold. 
Such a {\bf listen-before-talk}~\cite{lbt2006modeling, lbt2} allocation
strategy, in general, can be very conservative 
due to a combination of reasons: (i) the observed signal is actually an aggregate over all PUs in the same
band, (ii) path-loss need not be symmetric, (iii) SUs may want to
transmit at a much lower power than the PUs, and (iv) PURs may be a
distance away from the PUs.
\eat{
In this model, \textit{spectrum holes} are detected in an opportunistic 
fashion \cite{akyildiz2006next, devroye2008cognitive}, and spectrum is 
allocated to SUs in these area  with almost no restriction on the 
transmission power. E.g., in \cite{graphXi_2007} and \cite{graphTang06}, 
graph theory is used to find the spectrum holes and interference-free 
channel allocation.}
\blue{In static systems, a database approach~\cite{fcc35band}
can also be used by pre-computing
the SA function and storing in the databases; however, such an approach doesn't 
work in any dynamic situation, e.g., SUs \iab{at the tertiary} level, PUs changing powers,
path-loss changes due to real-time conditions.}
In a closely related work, ~\cite{gupta19} has developed
an interpolation-based spectrum allocation scheme that works by first
estimating the desired path-loss values based on signal strength
readings from deployed spectrum sensors. 
\eat{
The key challenge in estimating the desired path-loss values (SU to
the PURs) is that they are very different from the path-loss values
observed (``aggregated'' path-loss from the PUs) by the SS nodes.} 
In particular, they use inverse-distance weighted
and Ordinary Kriging interpolation schemes to estimate
the path-loss values. We compare our approach with theirs in \S\ref{sec:simulations}-\S\ref{sec:testbed}.
\eat{~\cite{curran2019procsa} develops a {\em secured}
protocol for the above scheme to preserve the privacy of involved entities.}

\para{Machine Learning (ML) Based Approaches.} 
To the best of our knowledge, there have been no
prior works that have used supervised learning to directly learn the SA function, 
especially as defined here. The closest
work is~\cite{azmat16}, which uses supervised learning to analyze
spectrum {\em occupancy} based on the sensed signal at the SU. 
\iam{Deep learning models have recently been used (see~\cite{cnn-3} for a
survey) to learn radio-propagation models and prediction; e.g., using CNNs~\cite{cnn-4}, for large-scale fading in 5G cellular networks~\cite{cnn-2}, using SegNet encoder-decoder model~\cite{cnn-1}, using SVMs~\cite{uccellari18}.
Path-loss models can then be 
used to allocate spectrum using Eqn.~\ref{eq:maxpower}---however, Eqn.~\ref{eq:maxpower} requires
knowledge of PU parameters, which may not be available.}
Moreover, the path-loss function
fundamentally encodes more information than the SA
function,\footnote{\mgtc{Note that the complete path-loss function 
is sufficient to estimate the SA function, but not vice-versa. Estimating
SA function from path-loss is straightforward in the \pset using
Eqn.~\ref{eq:maxpower}, while in \sset one can first estimate the PU parameters 
with reasonable accuracy 
from the path-loss function and SSs readings.}}
and thus, would likely require much more training.

\softpara{Reinforcement Learning (RL) Approaches.}
\mgtc{Reinforcement learning (RL) models have been used in slightly different
spectrum allocation settings wherein 
multiple agents are involved or competing with each other for spectrum resource
and they undertake a sequence of actions.}
E.g.,~\cite{fan2020,ghadimi17,zhang18,nasir21,chen2019reinforcement} have applied 
RL techniques for 
power control in multi-agent cellular networks, wherein the agents interact
with each other and cell towers to determine power allocation.
In addition, \mgtc{RL-based} spectrum allocation works in radio networks
(see~\cite{Morozs2015AcceleratingRL,wang2019survey} for surveys) have largely focused 
on channel assignment; in contrast, our work is focused on power allocation.
Zhang et al.~\cite{zhangDRL20} propose a deep RL algorithm where 
a deep neural network is used to help SUs obtain information 
about PUs' power policies.

\eat{
Multiple spectrum channels and power levels exist in their problem formulation.
Using Q-learning (an RL technique) algorithm, each D2D (SU) device is treated as an individual (independent) agent that is supposed to learn to find (for itself) the optimum channel and power level such that the overall throughput is maximized with no interference to PUs.}

\para{Multiple Channels and Other Objectives.} 
In this paper, we implicitly assume a single channel for the most part. 
Spectrum allocation for multiple channels can be done by using single-channel
techniques independently for each channel and then selecting one of the available channels based on some criteria (see \S\ref{sec:multiple_sus}).
For example, H.\ Wang et al.~\cite{wang10} picks a channel that maximizes the aggregate data rate of SUs, and X.\ Li et al.~\cite{li09} picks a channel that allows for minimum transmission power for a desired SU data rate.
Other works have addressed spectrum allocation with other optimization objectives. e.g., researchers have considered throughput maximization as an objective~\cite{salameh2011throughput,wang11} under various constraints such as
maximum allocated power~\cite{li09}, given QoS requirements~\cite{lee11}, etc. 
Fairness and energy efficiency are some other criteria considered~\cite{yu10,byun08,gao08}.

\subsection{Our Learning Approach}
\label{sub:contribution}

\para{Motivation \bleu{for Learning SA Function}.}
Our goal is to allocate spectrum efficiently to SUs in general settings, e.g., when PU parameters may not be available. To motivate and justify our learning based approach, we make the following remarks.
\underline{First}, we note that to
allocate spectrum near-optimally
to secondary users in general scenarios, we fundamentally need to have knowledge of the signal
path-loss function. In practice, however, even the best known path-loss models~\cite{chamberlin82,hata2000}
have unsatisfactory
accuracy, and conducting extensive surveys to gather path-loss values is infeasible and moreover, may
not even reflect real-time channel conditions. 
In absence of knowledge of a path-loss function, to allocate spectrum efficiently, we propose to
just learn the spectrum allocation function directly using supervised learning techniques.
\underline{Second}, in our context, an unsupervised approach is meaningless as unlabelled samples have minimal information (actually, zero information in the \pset), and \mgtc{as explained
in \S\ref{sec:deep_learning}, a reinforcement-learning approach is also not suitable for our setting.}
\underline{Third}, learning the path-loss function first and then using Eqn.~\ref{eq:maxpower} to allocate 
spectrum is certainly a feasible approach -- but, since the path-loss function fundamentally encodes more information than the SA function, it would likely require much more training (note that
the SA function depends only on the most restrictive of the PURs). 
\bleu{
\underline{Finally,}  non-trivial parameters such as weather, terrain and obstacles, PU transmitters being directional, etc., can be relatively easily incorporated  in a learning approach (see \S\ref{sec:multiple_sus}), while they would require more sophisticated modelling techniques and algorithms to be incorporated 
in non-learning approaches.}

\tmagenta
Based on the above observations and insights, we propose to learn the SA function directly from training examples. In general,
our goal is to learn the SA function accurately with minimal training.
Below, we discuss the inputs/features of our learning models, and 
the gathering/generation of training samples.

\para{Inputs/Features of the SA Function \A.} For most of the discussion in this paper,
we focus on spectrum allocation to a {\em single/first} SU in a given area; multiple
or subsequent SUs can be handled similarly as discussed later 
in~\S\ref{sec:multiple_sus}.
In the simplest of settings, the PUs' information/parameters (location and transmit power) do not change over time, in which case the SA function can be simply represented as a function of just the SU's location, 
as the fixed PU parameters will be 
automatically captured within the learned model.
In this work, we focus on the more general settings wherein the PUs' information (location and power) may change across training and evaluation samples. 
For our two settings, viz., \pset and \sset, 
inputs/features of our SA function are as follows.

\begin{enumerate}
\item \pset. In this setting, the PUs' parameters are available for
spectrum allocation determination. Here, the inputs to the SA function are 
PU parameters \PUParam  and location \loc  of the requesting SU, and thus,
the SA function can be represented as a real-value function 
$\mathcal{A}(\PUParam, \loc)$.
        
\item \sset.
In this setting, PUs' parameters may not be available, and, 
as mentioned before, we use spectrum sensors (SSs) to gather received power
and use sensors' readings to determine spectrum power allocation. 
Here, the inputs to the SA function are (i) SSs parameters $\Sensor$,
and (ii) location \loc of the requesting SU. Thus, the SA function is
can be represented as $\mathcal{A}(\Sensor, \loc)$. 
For each SS, its parameters may include its location and aggregate received power from the PUs, and in general, may also include the mean and variance of the Gaussian distribution of the received power. 
\end{enumerate}
\tblack

\para{Gathering and Labeling Training Samples.} Note that gathering a training sample for \A entails gathering feature values and determining its ``label''---in our context, \mgtc{for a given feature vector (SU location and PUs/SSs parameters), the label is the maximum power that can be allocated to the SU without causing harmful wireless interference at {\em any} of the PURs. 
In the \pset, the features are the available PU parameters.
In the \sset, for gathering training samples, we need to deploy SSs and gather received powers (see \S\ref{sec:testbed}) for collecting sample features. 
In either setting, to label the sample, we need to estimate \iab{the} maximum allowable power
at a given SU's location; this entails simulating PURs and determining the maximum 
SU transmission power that allows PURs to receive the PU signal (i.e., ensures that
the signal-to-noise ratio is above a desired constant).
To estimate the maximum SU power allowed, we can do a binary search on SU power, 
as done in our tested setup described in \S\ref{sec:testbed}.
To circumvent \iab{the} collection of noisy samples, we can throw out samples that cannot be reproduced.
We acknowledge that the training process can incur a substantial cost, but can be automated using drones for entities. More importantly, training is done 
only one-time, and thus, some amount of training cost is tolerable.}

\section{CNN-based Deep Learning Approach} 
\label{sec:deep_learning}

In this section, we motivate \iab{the} convolutional neural network's (CNN) suitability 
for our context, and design an efficient CNN architecture and associated techniques for our problem. 
\tmagenta
\para{Motivation for Using CNNs for SA Function.}
We observe that SA function can be looked upon as an 
image regression function, with 
the inputs to the SA function (SU location, PUs/SSs parameters, etc.) 
represented as an image
and the regression value representing the optimal allocation power. 
Framing SA function as an image regression function allows us to 
leverage known advanced image regression models.
In particular, CNNs have been very successful in image classification or 
regression tasks or in capturing patterns/objects in images, 
because CNNs are able to exploit the spatial structure in images
via \iab{the} use of learnable spatially-localized filters (or kernels) in its 
convolution layers.
In our context, the spatial nature of the SA problem 
(i.e., entities deployed over a geographic area) means that 
the input to the SA problem can be represented  as a 2D image 
with sufficient accuracy.
Such an input representation allows us to use CNNs to 
learn the SA function effectively. 
We corroborate our above intuition about the suitability of CNNs to our problem via extensive evaluations in~\S\ref{sec:simulations}.
For the general case of 
multiple SUs, we augment our CNN model with Recurrent-Neural Networks (RNNs) in
\S\ref{sec:multiple_sus}.
\tblack

Radio propagation modeling and prediction using CNNs cnn-4
overview cnn-3
large-scale fading in 5G cnn-2




\softpara{Challenges.}
However, there are significant challenges that need to be addressed, 
to make \blue{CNN} a viable and efficient
approach to learn the SA function.
These include the pre-processing  of samples into "images" to feed as input to a \blue{CNN}
model, creating an efficient CNN architecture,
ensuring minimal false positives, handling multi-path fading effects, 
minimizing training \iab{costs}, etc.
We discuss these challenges in the following subsections. 

\para{Other Machine Learning Models.}
We believe that CNNs are best suited to model the (single SU) SA function.
However, other machine learning models can also be used to learn the 
SA function---in particular, we also evaluate
neural networks (NNs) and space-vector machines (SVMs) approaches in \S\ref{sec:eval-single}. 

\mgtc{
However, we note that reinforcement learning (RL) approach 
is not suitable for learning the spectrum power allocation function, 
as defined in this work, for the following reasons.
Learning our SA function is fundamentally a supervised learning 
problem---since our learning goal is to approximate a function using 
{\em labeled} training examples (which can be gathered, as 
discussed before).
In contrast, in RL settings, an agent learns  
a policy about which action to take on each state 
so as to maximize the cumulative reward; 
the learning of such a policy is driven by 
(i) the rewards given by the environment based on the current/next state(s) 
and/or the action taken,
and (ii) the Markov-decision process (MDP) over the system states 
and actions that \iab{represent} how the system transitions through states potentially
influenced by actions.}
As there is no underlying MDP governing our spectrum allocation 
setting, the RL approach is not suitable for our 
context.\footnote{If we use the RL technique in our setting by considering
actions as power allocations, we'll need to provide training examples for {\em every} possible system state, due to \iab{the} lack of an underlying MDP, making the approach infeasible. Note that in a setting with no underlying MDP, the RL approach learns the policy independently for each state.}

\subsection{\sname: Shallow CNN Model}
\label{sec:basic-cnn}

In this subsection, we discuss our basic CNN architecture and approach, 
which we refer to as \sname, as it 
has a small number of layers. In the next subsection, we will extend this approach to the 
\name approach that uses a much deeper CNN architecture with a larger number of layers. We
start with discussing our strategy to pre-process training samples into images.

\para{Pre-Processing Training Samples to Images.}
The first challenge in applying CNNs to our context effectively is to 
transform each training sample to an “image” for input to the CNN model, in a way that
it is most conducive to efficient learning.
Representing the entities (PUs, SSs, SUs) as objects in a 2D image
is a natural choice. 
In particular, we could represent each entity type with a different color or shape, or more specifically, represent each entity by a disk of a certain color with a radius based on the transmit/received power.
Our choice of image representation is tantamount to \textit{feature engineering}~\cite{bengio2017deep}, and can have a significant impact on the training
cost.
Below, we discuss the choice made in our model design.
\rdtc{First, we assume 
that the PURs are distributed similarly across all PUs; 
more formally, we assume that the distribution of PURs around its PU $P$ is a function of $P$'s parameters (location and power).  
Under the above assumption, we do not need to represent PURs of any PU 
in the input image---as the distribution of PURs can be learned by the 
model from each PU's parameters.
E.g., if for each PU $P$, its PURs are
distributed uniformly within a fixed radius around $P$ (or within a radius proportional to $P$'s power), then we don't need to represent PURs in the input image. 
}

\begin{figure}[h]
    \includegraphics[width=0.35\textwidth, center]{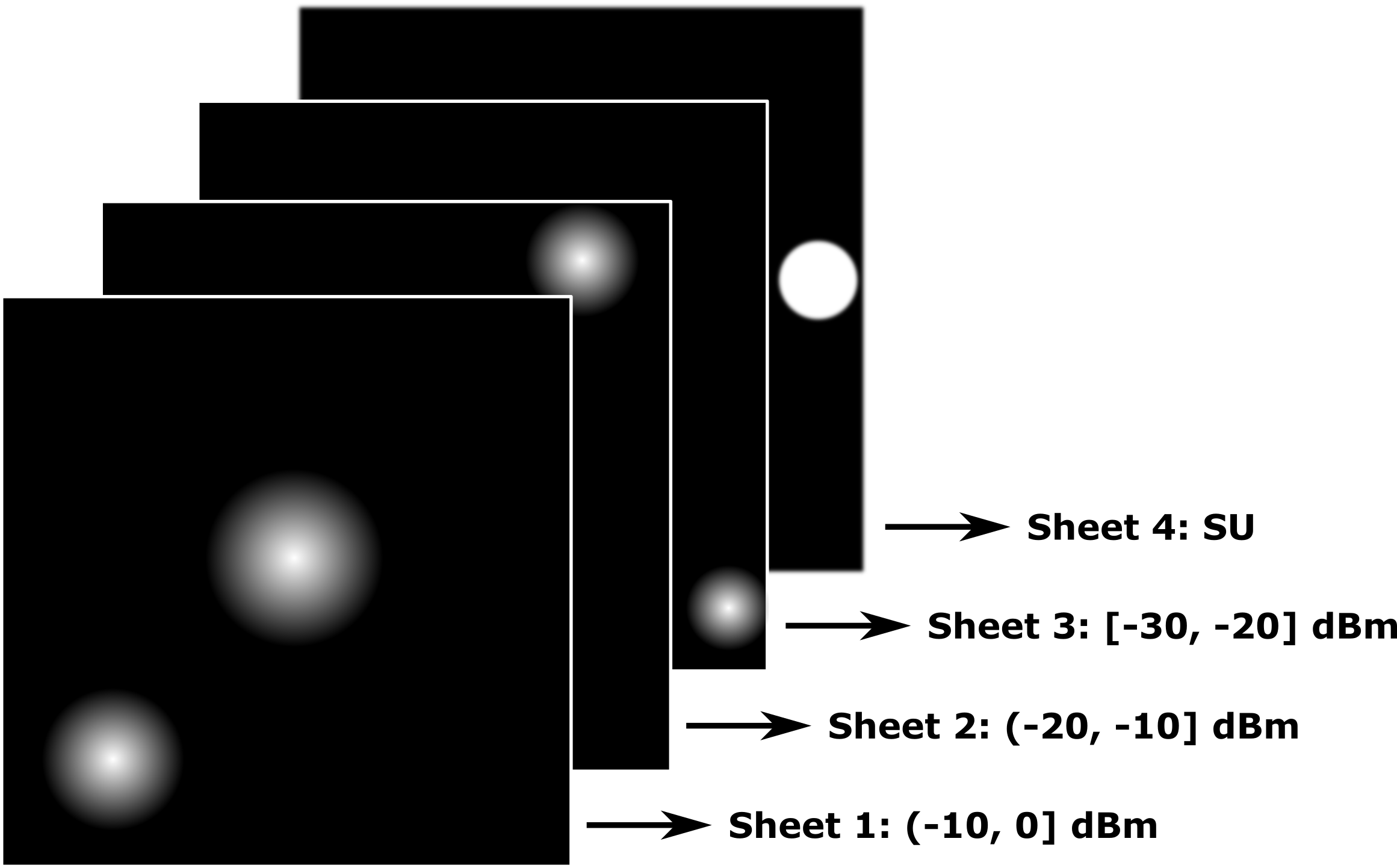} 
    \caption{Image representation of a sample for input to the \blue{\name} model. Here, there are four PUs (2 in the first sheet, and 1 in the other two sheets) and one SU (in the fourth sheet). Each sheet for the PUs corresponds to a range of PU's transmit power; e.g., PUs whose transmit power is in the range -10 to 0 dBm are placed in the first sheet. \bluee{When representing SSs in \sset, we place SSs in the sheets based on their locations.}} 
    \label{fig:cnn_input}
\end{figure}
\softpara{\pset: Representing PUs in Multiple Image ``Sheets.''} 
Note that just using shapes or colors for different
entities is \iab{insufficient,} as we also need to represent \iab{the} 
transmit/received powers. 
Just using radius to represent powers is not viable either,
as we may start getting intersections between shapes. 
Thus, we compose the input image of a certain number of ``sheets'' 
(see Figure~\ref{fig:cnn_input}). Then, 
we divide the expected transmit-power range into ranges 
\bluee{of about 5-10dBm} and assign each PU to the appropriate image sheet depending on its transmit power. 
Within each sheet, 
we then use disks to represent each PU, with the brightness of the center 
pixel as well as the radius of \iab{the} disk proportional to the transmit power. In addition, to give more importance to the center (which represents
the true location) and to suggest signal attenuation away from the center, we decrease the brightness of the pixels away from the center in a logarithmic manner. If there is an intersection between two objects in the same sheet, 
we aggregate the intensity of the common pixels.
Unlike normal images, which are composed of three sheets corresponding to red, green, and blue colors, we
may use more than three sheets.

\softpara{\sset: Representing Spectrum Sensors' (SSs) Readings.}
SSs can be represented \iab{similarly} to PUs with the size of their disk proportional to the {\em received} (rather than transmit) power. \bluee{However, unlike PUs, we place SSs among the sheets based on their location rather than received powers, e.g., SSs from certain grids were always placed on the first sheet irrespective of their received-power readings. In our evaluations, we observed that placing SSs over sheets
based on locations improved the performance of our models significantly compared to placing the SSs based on received powers.}
\eat{Note that, in our context, we don't 
need to represent PUs and SSs together, but, if information from both PUs
and SSs is available, they can be represented in different sheets. We choose
the radius of the SSs to be smaller than that of PUs as the density of SSs
in an area is expected to be much more than the PUs to avoid many
intersections among the disks.}

\softpara{Representing SUs.}
\blue{We use a separate sheet to represent SU(s), instead of representing them with a different shape/color, to facilitate potentially more efficient training.}

\para{\sname CNN Model Architecture.}
To design a \blue{CNN} architecture to learn the SA function efficiently, 
we need to carefully determine the various model parameters.
For our \sname model, we choose these parameters as follows. 
{\bf (i)} Number of convolution layers in a 
\blue{CNN} model plays an important role in training cost as well as model accuracy.
In general, a deeper model (i.e., with more number of convolution layers)
performs better than a shallow one but incurs more training \iab{costs.}
In our context, only a few thousand training samples \iab{are} feasible to gather, 
we use 5 convolution layers, and 3 fully-connected layers as in a neural-network 
architecture.
{\bf (ii)} {\em Filter Size:} 
To be able to "detect" small-radius PUs of small power values, we use 
a $3\times3$ filter in the convolution layers. The number of filters
is small for the initial layers, and then increases with
the "depth"; this facilitates the detection of more and more complicated
features in the deeper layers.
{\bf (iii)} {\em Activation Function:} 
In our context, as the final output of the model is a real number, we chose a linear activation function for the last network layer with only one neuron.
For other layers, we use Rectified Linear Units (ReLU)~\cite{nair2010rectified} activation function as its non-linearity enables the model to learn a 
more complex function \blue{faster}.

\begin{figure*}[t]
\centering
\begin{minipage}[t]{1\textwidth}
    \includegraphics[width=\textwidth, center]{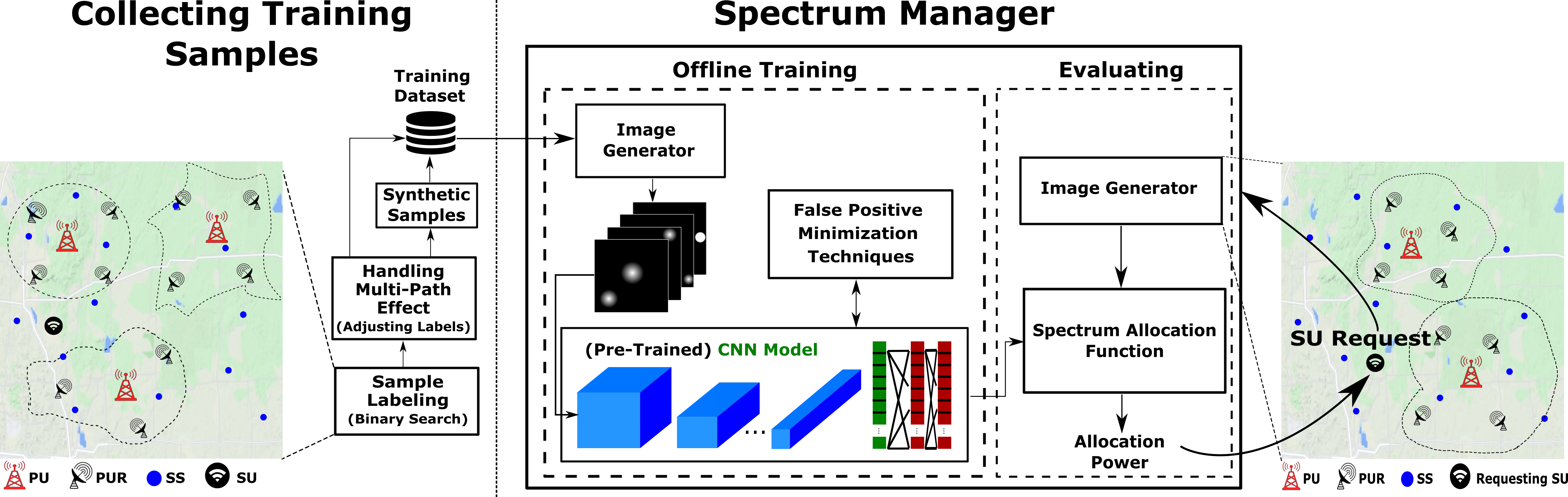}
      \caption{CNN-based Spectrum Allocation System. For the CNN Model component, \sname uses a shallow CNN model without any pre-training, while \name uses a deep pre-trained CNN model obtained from Fig.~\ref{fig:deep_figure}. In both
      cases, after any pre-training, the (field) training samples are gathered, converted into images,
      and then used to (further) train the \blue{CNN} model. The learned model is then used to
      allocate spectrum to requesting SUs.}
   \label{fig:cnn_system}
\end{minipage}
\hspace{0.1in}
\end{figure*}

The overall CNN-based spectrum allocation system is shown in Fig.~\ref{fig:cnn_system},
including components discussed in the following subsections. 

\subsection{\name: Pre-Trained Deep CNN Model}
\label{sec:deep-cnn}

\begin{figure*}[t]
\centering
\begin{minipage}[t]{1\textwidth}
    \includegraphics[width=\textwidth, center]{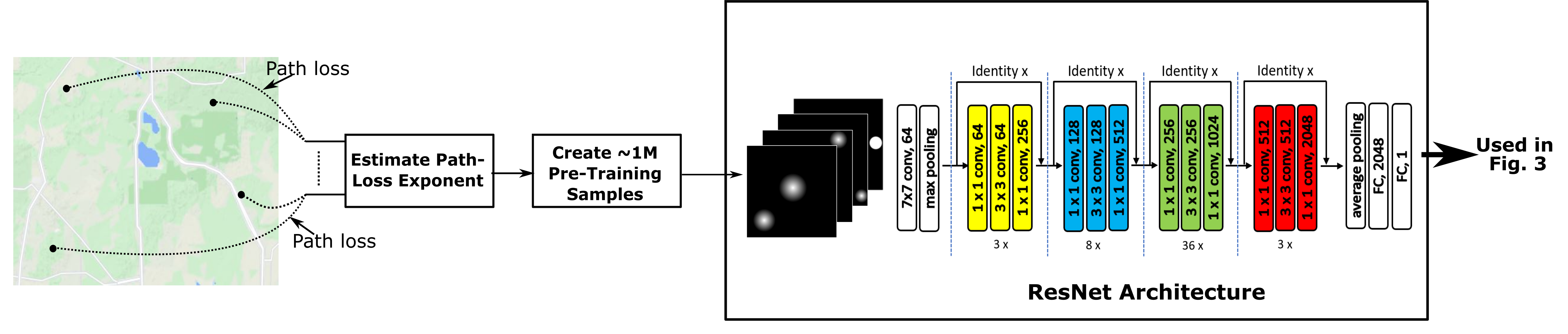}
      \caption{\name Pre-Training Process. Here, we generate a large number ($\approx$ 1M) of images assuming a log-normal propagation model based on an estimate path-loss exponent, and pre-train a deep CNN model (we used ResNet) using these generated images. The pre-trained model is then 
      used in Fig.~\ref{fig:cnn_system} for further training using a smaller number of field training samples.}
   \label{fig:deep_figure}
\end{minipage}
\hspace{0.1in}
\end{figure*}

It is well understood that deeper neural networks, i.e., neural networks with more layers,
and in particular, deeper CNN models can yield much better performance with their ability to 
learn more complex functions, with sufficient training data. 
However, there are two major challenges with using deep models. First, as the number of layers
increase, so does the number of parameters to learn---which in turn requires
a much larger
number of training samples.
Second, a deeper model is more likely to learn a function that overfits the training data, if 
the training data is not sufficient. 
The perfect solution is to use a deep network but also have a sufficiently large
training set. However, in our context, it is infeasible to gather more than a few thousand 
samples as the deployment and/or \iab{labeling} costs are high, while a very deep
(e.g., with 20-100 layers) network may require close to a million samples to train properly~\cite{vgg,resnet,xception}.

\para{Using Pre-Trained Models.}
Our approach to address the above challenge is to use a {\em pre-trained} deep model, i.e., a
deep model that has already been trained with readily-available images, which may not represent
spectrum allocation samples. Such a strategy has been often used in computer vision applications with
great success. Similarly, we could also use pre-trained deep well-known 
models such as VGG~\cite{vgg},
ResNet~\cite{resnet}, Xception~\cite{xception} 
which have been trained with around 
1 million daily-life images, and then train them further
with a few \iab{thousand} spectrum allocation training samples, as in the previous subsection.

\softpara{\name Approach.}
One unique aspect of our context is that the pre-training samples or images can be easily 
synthesized based on an assumed propagation model. Thus, to obtain better performance than
the above models \iab{are} pre-trained with daily-life images,
we can pre-train a deep architecture \bluee{(we use ResNet~\cite{resnet}\footnote{We evaluated VGG-based \name too but it was easily outperformed by the ResNet-based \name architecture
especially in the \sset.})}
with generated spectrum allocation images based on an assumed log-normal
propagation model with an "appropriate" path-loss exponent. \bluee{(Results in Fig.~\ref{fig:multi_regions_pretrained}(b) of~\S\ref{sec:eval-single} validate the above pre-training approach.)}
The path-loss exponent $\alpha$
can be derived by gathering a few (say 100-200) path-loss samples, and determining the best
$\alpha$ that minimizes the error between the actual samples and those from
the log-normal model. See Fig.~\ref{fig:deep_figure}.
Then, as before, we train such a pre-trained model further with a few thousands of 
SA training samples as shown in Fig.~\ref{fig:cnn_system}. We refer to this
overall approach as \name.


\subsection{Minimizing False Positive Error}
\label{sub:interference}

Note that we should rarely allocate power 
higher than that determined by Eqn.~\ref{eq:maxpower}, as it 
would cause harmful interference to some PURs. 
Ideally, we would like to minimize such cases, which we call {\em false positives}, drastically---perhaps, at the cost of higher
number of (and/or errors in) the false negative cases. 
We address this issue by a combination of two strategies. First, we choose the hyper-parameters
of the \blue{CNN} model in a way that minimizes the number of false positives; the methodology here is
largely trial and error, due to \iab{a} lack of techniques to systematically search for hyper-parameter
values. 
Second, we use asymmetry in the training loss function as follows. Essentially, we 
change the training loss function $\textbf{J}(\boldsymbol{\theta})$ such that false positive samples (SU requests that get higher power than the maximum) are penalized more drastically than the other requests. 
More formally, we define
the training loss function as:
\begin{equation}\label{eq:AsymLoss}
    \textbf{J}(\boldsymbol{\theta}) = \frac{1}{m} \left(\afp\sum_{\hat{y_i} > y_i}{\textit{l}(\hat{y_i}, y_i, \boldsymbol{\theta})} + 
    \afn\sum_{\hat{y_i} \leq y_i}{\textit{l}(\hat{y_i}, y_i, \boldsymbol{\theta})}
    \right)
\end{equation}
Above, \afp and \afn are coefficient weights for the false positives and false negatives respectively, $y_i$ is the "ground truth", $\hat{y_i}$ is the predicted value, $\boldsymbol{\theta}$ is the internal set of the parameters being learned, and \textit{l}(.) defines the error for a single sample.
To minimize false positives, we can choose a much higher \afp than 
\afn.
\begin{figure}[t]
\captionsetup[subfloat]{farskip=0.5pt,captionskip=0.5pt}
\centering
    {
        \includegraphics[width=0.9\linewidth]{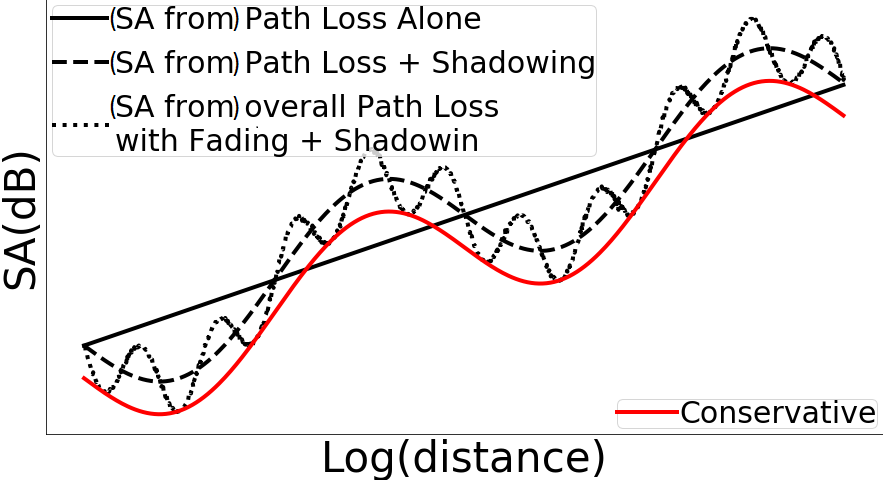} 
        \vspace{0.2cm}
    }
    {
        \hspace{-0.1cm}
        \includegraphics[width=0.9\linewidth]{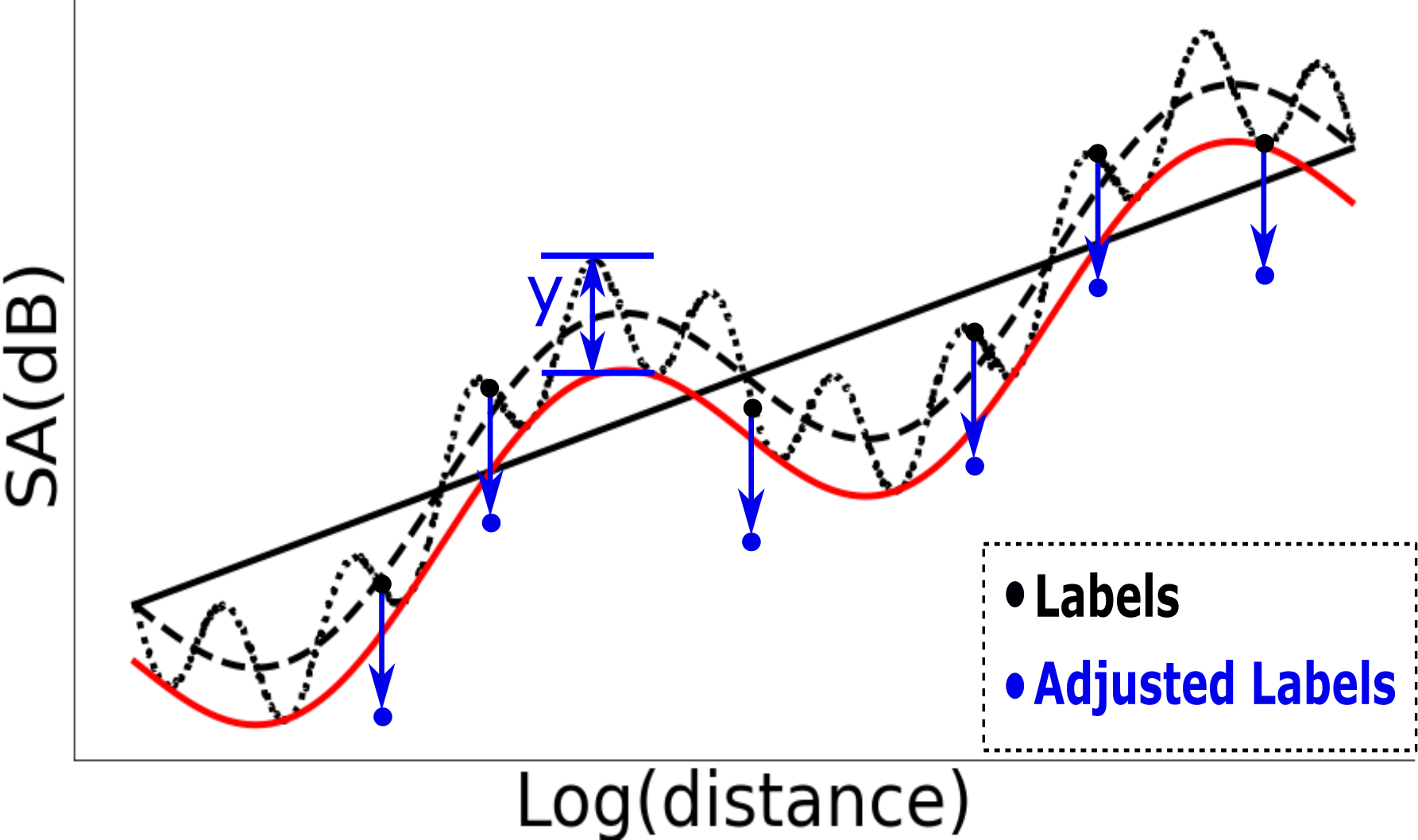} 
    }
 \caption{(a) Spectrum allocation function, when there is a single PU and a single SU, due 
 to path loss with shadowing and multi-path fading effects. Note that the path-loss function
 has a similar trend. The red plot is a conservative spectrum allocation, based on a 
 similar conservative path-loss function. (b) Modifying labels of training samples to drive the model towards learning a simpler and more conservative spectrum allocation function.} 
 \label{fig:path_sa}
\end{figure}
\subsection{Handling Multi-Path Effect}
\label{sub:multipath}

Wireless signal attenuation is a result of three mutually independent, multiplicative propagation phenomena~\cite{shankar2017fading},
viz., large-scale signal attenuation, medium-scale shadowing, and small-scale multipath fading. 
The shadowing effect can cause overall path loss to vary over 10s to 100s of meters, and the fading effect can cause an effect over a few wavelengths.
See Fig.~\ref{fig:path_sa} (a).
The overall effect of these phenomena in our context is that learning the SA function could require a much higher number of
training samples---to sufficiently and accurately capture the small to
medium scale fluctuations. 
This challenge is fortunately mitigated by the
fact that the small-scale fluctuations in the path-loss function are
unlikely to fully manifest in the SA function, as the SA function
depends only on the most restrictive of the SU-PUR path losses.

In our context, one way to address the above challenge is to learn instead of a more ``conservative'' SA function that may be easier to learn but may
allocate sub-optimal power. E.g., the conservative SA may be based on
a {\em lower-bound} approximation of the path-loss function. See the red curve
in Fig.~\ref{fig:path_sa} (a).
We note that the technique in the previous section is also 
driven towards learning a more conservative function via an appropriate training loss
function. However, the model inaccuracy resulting from the multi-path effect is
fundamentally due to a lack of sufficient training samples
needed to learn a function with high spatial resolution/variability;
thus, we can't fix the model inaccuracy due to the multi-path effect
by merely changing the loss function, model, or its training process. 
Our strategy to learn a more conservative function that addresses the
multi-path effect challenge is to actually modify the labels
of the training samples appropriately, as discussed below. 

\para{Learning a Conservative SA Function by Modifying Training Sample Labels.}
To ensure few false positives in \iab{the} face of the small-scale multi-path 
\iab{the} fading effect, we modify the training samples by lowering their labels from
the optimal allocated power to smaller values so that the modified training
samples represent a "conservative" spectrum allocation function without the 
small-scale effects.
In particular, let's consider a training sample--corresponding to a certain set of 
PUs with given transmit powers, PURs, and SU $\sii$ with allocated power $\Pi$ which
satisfies Eqn.~\ref{eq:maxpower}.\footnote{However, note that $\Pi$ is not computed {\em using} Eqn.~\ref{eq:maxpower}, as many terms in it are unavailable.}
(Our discussion in this subsection applies to both settings, viz., \pset and \sset.)
Let $l_i$ be the location of the SU \sii. 
Now, let PUR $\rj$ be the PUR that results in \sii's allocated power, i.e., let
$j = \argmin_{j}\frac{\tau_j - I_j}{\rho(l_i, l_j)}$ from Eqn.~\ref{eq:maxpower}. 
In the following discussion, we vary $l_i$ while keeping everything else constant--thus, we can look at $j$, $\rj$, and allocated power $\Pi$ as functions of $l_i$.
Now, in a sufficiently small neighborhood $N$ of $\sii$, 
the PUR $\rj(l_i)$ as computed above and thus $\Pi(l_i)$  must remain fixed for  $l_i \in N$.
Moreover, $\Pi(l_i)$ is also inversely \iab{proportional} to $\rho(l_i, l_j)$ with $l_j$ fixed, and more
importantly, variation of $\Pi(l_i)$ within $N$ is largely dominated by the small-scale effect of
the path-loss function. Let $y = \max_{l_i \in N} \Pi(l_i) - \min_{l_i \in N} \Pi(l_i)$, which in 
some sense is an estimate of the "amplitude" of the small-scale effect. 
We now use the value of $y$ to lower the allocated power label of the training samples. See Fig.~\ref{fig:path_sa}(b). The modified samples essentially correspond to a conservative 
spectrum allocation function, without the small-scale effect. 
Note that the value $y$ depends on the small-scale effect of the terrain. If we assume the effect
to be uniform across the given region, then the same value of $y$ can be used across all the 
training samples---else, we compute $y$ in each subarea.

\subsection{Synthetic Samples to Improve Performance} 
\label{sub:cost}

One way to improve performance, i.e., to aid the model in extracting the most information 
from the given training samples,
is to create additional \textit{synthetic} samples from the training samples. 
In effect, the new synthetic samples incorporate the domain knowledge used
in creating them.
In general, from a given sample $\{X, y\}$ where $X$ is the set of features (PU parameters or SS readings, and the requesting SU's location) and $y$ is the label (allocated power), we create synthetic samples of the type $\{X', y\}$ where $X'$ is another set of features that yields (approximately) the same label $y$. 
Below, we discuss the generation of synthetic samples for our two settings, \mgtc{and evaluate the improvement from these strategies in \S\ref{sec:eval-single}.}

\para{PU-Setting.} 
Consider a sample $\{X, y\}$ where $X$ includes the PU parameters. 
For this sample, let $P$ be the PU for the PUR \rj that determines the SU's optimal
power allocation, as in \iab{the} previous subsection. 
Let $\mathcal{P}$ be a set of PUs that are sufficiently far away
from $P$.
Note that $P$ is determined when labeling the given sample, and  $\mathcal{P}$ can be determined from $P$ and the available PU parameters. 
Now, note that decreasing the power of PUs in $\mathcal{P}$ should not change the optimal power
allocated to SU. 
Thus, we can create additional synthetic samples by considering $X'$ which differs
from $X$ in that the transmit powers of PUs in $\mathcal{P}$ is lower than that in 
$X$. 

\para{SS-Setting.}
In the setting, wherein the features are composed of sensor readings, we generate synthetic
samples by determining sensor readings at additional locations via interpolation techniques. 
In particular, based on the log-distance-like behavior of signal attenuation, we use a 
slight modification of the traditional IDW technique. More formally, for a known set of 
sensor readings $(p_i, l_i)$ where $p_i$ is
the received power at location $l_i$, the interpolated value $q$ at a new location $l$ is given
by: $q=\frac{\sum_i w_i p_i}{\sum_i w_i}$, 
where $w_i$ is the weight defined as $w_i = \frac{1}{\log_{10}(d(l_i, l))}$.
Based on the above interpolation scheme, for a given sample $(X,y)$ where $X$ is the set
of real sensor readings, we can \iab{create} synthetic samples of the type $(X', y)$ where $X'$ consists of some readings from $X$ and some interpolated readings.

\para{Rotated Images.} In addition to \iab{the} above, we also synthesize additional samples by just "rotating" the images of the original samples; such a strategy is bound to be useful in regions where the propagation model is largely dependent on the distance. 
In our evaluations, we rotate the given images by 90 or 180 degrees. 

\subsection{Multiple SUs}
\label{sec:multiple_sus}

Till now, we have only considered spectrum allocation to the initial single SU.
Handling {\em subsequent} SUs in the same manner as the initial SU above requires
allocated power to the active SUs.
We start with discussing how to handle multiple SUs {\em one at a time}, and
then discuss handling multiple SUs simultaneously.

\para{Handling SUs One at a Time.}
To handle subsequent SUs, we can augment the list of features to include the active SUs' parameters and train the model accordingly. 
This approach is viable, especially in the context of our \name model,
since active SUs with their parameters (location and allocated power) can 
easily be represented in the single input sheet dedicated to SUs. The only
change we need to make is that we must now represent SU's allocated power
too; we can represent them with a disk similar to PUs, i.e., with the
center's brightness and radius proportional to the allocated power. 
If multiple SU requests arrive \textit{together}, then to use the above
approach, we need to decide on an order in which to handle the requests.
One approach is to handle them in a greedy order as follows; we call
 this approach  \nameg. 
We divide the whole area into non-overlapping subareas and assign each subarea $X$ 
a weighted score equal to the total aggregated power transmitting from all the PUs and 
active SUs in $X$ or its adjoining cells.
Then, we pick the SU requests in the ascending order of the weight of the subarea 
they belong to.

\begin{figure}[t]
\centering
\begin{minipage}[t]{1\linewidth}
    \includegraphics[width=0.9\linewidth, center]{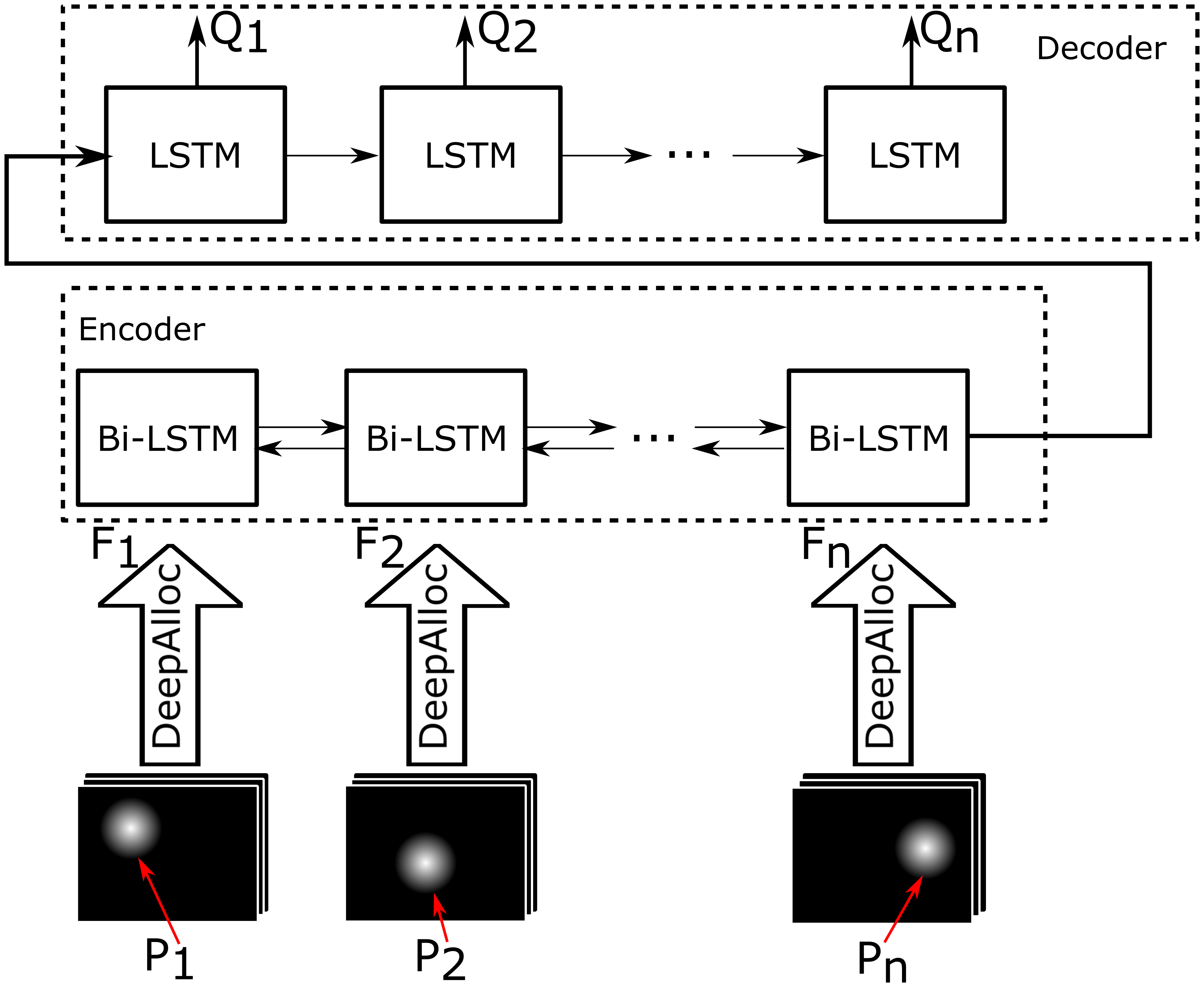} 
      \caption{RNN-based Architecture for Simultaneous Allocation to Concurrent SUs. The input to the encoder is the convolution output $F_i$ of \name for each SU $S_i$, and the output of the  model is the SU power allocations $\{Q_i\}$. The input to \name  
      for each $S_i$ is a set of sheets representing $S_i$'s location $l_i$
      and independent power $P_i$ and PUs/SSs information.}
   \label{fig:rnn_system}
\end{minipage}
\hspace{0.1in}
\end{figure} 

\para{Handling Multiple SUs Simultaneously.}
Handling SUs one at a time, though easier and simpler, can lead to unfair and/or
far-from-optimal (in terms of say, {\em total} power allocated) allocation.
Thus, we now discuss how to extend our deep-learning models to facilitate
simultaneous allocation of powers to multiple SUs.
Recall that our \name model 
is trained to predict the maximum power that can be allocated to a single SU.
\tmagenta
To allocate powers to multiple SUs $\{S_1, S_2, \ldots, S_n\}$ simultaneously, 
we need to essentially learn a function that maps 
the list of SUs' locations $\langle l_1, l_2, \ldots, l_n \rangle$ 
to the list of simultaneous 
allocated powers $\langle Q_1, Q_2, \ldots, Q_n \rangle$. 
To learn the above function, we use two different learning approaches.

\softpara{(1) \namenn.} The simplest approach is to use a traditional 
neural-network (NN),
with inputs as the list of SUs' locations $\langle l_1, l_2, \ldots, l_n \rangle$ 
and other available PU or SS parameters, and the output as the list of allocated simultaneous 
powers $\langle Q_1, Q_2, \ldots, Q_n \rangle$.  To leverage our developed 
single-SU model \name and facilitate more efficient training, 
instead of passing the PU or SS parameters, we pass the set of ``independently allocated''
powers $\{P_1, P_2, \ldots, P_n\}$ where $P_i$ is the power allocated to $S_i$ 
assuming no other SU $S_j (j \neq i)$ is active.
Thus, the input to the NN is the list $\langle (P_1, l_1), (P_2, l_2), \ldots, (P_n, l_n) \rangle$, with the output being the list of simultaneous 
power allocations $\langle Q_1, Q_2, \ldots, Q_n \rangle$. 
The NN is trained using
training samples with labels determined as described below. In our simulations,
we used a 4-layer NN with 128 neurons in each layer and a dropout of 80\% before 
the last layer.


\softpara{(2) \namernn.}
As mentioned above, we need to learn a model that maps a list/sequence of SUs' locations
to a list/sequence of allocated powers. In recent years, the \textit{Seq2Seq} models 
have achieved a lot of success in learning tasks that \iab{involve} mapping a sequence of words to another sequence of words, e.g, in machine translation, 
text summarization, image captioning, etc. In particular,  Google Translate 
uses \textit{Seq2Seq} model to translate a paragraph from 
one language to another.
The \textit{Seq2Seq} model comprises of an encoder to encode the input sequence, 
and a decoder to map the encoded data to a target sequence. The encoder
essentially creates a sequence of hidden state vectors each of which 
depends upon the previous input and state vector. Similarly, the decoder uses the
created hidden state vectors with appropriate attention/weighting schemes to generate 
a target sequence.

As our multi-SU spectrum allocation problem entails mapping a sequence of SUs' 
information to SUs' power allocations, we develop a \textit{Seq2Seq} model with appropriate inputs, outputs, and
internal components as follows.
\tblack
For our purposes, we propose to use an RNN-based Seq2Seq model, 
which we call \namernn. In 
particular, we implement the encoder using a bi-directional long short-term 
memory (LSTM) model and the decoder as a uni-directional LSTM. See Fig.~\ref{fig:rnn_system}. 
\tmagenta
We use bi-directional LSTM as the encoder to facilitate interpretation of the input set of SUs as {\em unordered} to signify {\em simultaneous} allocation.
In addition, rather than passing $\langle (P_1, l_1), (P_2, l_2), \ldots, (P_n, l_n) \rangle$
as the input to the RNN model as we did in the above NN approach,
we pass the set of feature-sets $\{F_i\}$ as input where $F_i$ is as follows (see below for the motivation).
\tblack
The feature-set $F_i$ (for a given $S_i$) is the output of the average-polling 
layer (Fig.~\ref{fig:deep_figure}) of the \name model,
when the input (represented in sheets) is SU $S_i$'s location $l_i$
and independent power $P_i$ and given PUs/SSs parameters.  See Fig.~\ref{fig:rnn_system}.
The above idea of passing $F_i$'s to the RNN allows us to encode more 
information in the input to the RNN, rather than just the power $P_i$'s.
The final output of the decoder represents the powers allocated to the SUs together.

\softpara{Determining Training Sample Labels (\bina).}
To train the above models, we need a method to label the training samples; in particular,
we need to design an algorithm to determine \iab{the} optimal powers to allocate to multiple 
concurrent SUs. Our implicit performance metric is the {\em total} power allocated 
to the given set of SUs.
Here, in the training phase, even though we can assume knowing whether a certain power
allocation to multiple SUs causes interference to any PUR, to determine optimal power 
allocation to multiple SUs is still non-trivial. 
Since we couldn't design an optimal  algorithm, we \iab{used} the following
\bina heuristic which performed well in practice. 
The \bina heuristic is a  
binary-search-like iterative algorithm as follows.
At any iteration, each SU $i$ is assigned a {\em range} $[l_i, u_i]$ 
of powers such that if {\em all} SUs are allocated the lower-end $l_i$ of their ranges
then there is no interference caused to any of the PURs.
The algorithm iteratively reduces (in half) the range of one of the SUs, till
the range of each SU is smaller than \iab{the} threshold. 
In particular, in each iteration, the algorithm picks the SU $i$ with the largest range $(u_i-l_i)$,
and either moves the lower-bound $l_i$ or upper-bound $u_i$ to $(u_i+l_i)/2$ depending
on whether $(u_i+l_i)/2$ causes interference to a PUR.




\para{Multiple Channels; Request Duration.} 
Till now, we have implicitly assumed a single channel. 
To handle multiple channels, we can extend our schemes easily.
If the signal propagation characteristics are different for different channels,
then we can create a separate model for each channel.
Then, for a single-SU request, we can pick the channel that 
allows for maximum power allocation.  
\mgtc{We can also modify the above multi-SU approaches, for the multi-SU scenario,
as follows.}
The \nameg approach extends naturally. 
For \namenn and \namernn, we use a simple scheme of uniformly (and randomly) partitioning the given set of SUs into multiple sets and assign a channel to each set; incorporating channel assignment in the model would require much more training, and thus not considered for simplicity.
\mgtc{Finally, to handle requests for a specified duration, 
whenever a currently active SU becomes inactive (i.e., at the end of its authorized duration), we reallocate other active SUs' allocations.} 

\para{Other Generalization: Non-Isotropic Users, Weather, Obstacles.}
\bleu{We now briefly discuss how our approach can be generalized to handle more general and sophisticated scenarios. 
We start with discussing how to handle non-isotropic PUs or SUs.
In the \pset, the non-isotropic PUs can be represented by
an appropriate cone (instead of a disk). In the \sset, the SS readings would 
naturally incorporate the impact of non-isotropic PUs---and hence, there are no
changes to the approach.
To handle changes in propagation models due to different weather conditions, 
we can train different models for different weather conditions (e.g., rainy, 
summer, winter, etc). To incorporate more fine-grained weather conditions, 
we can use a separate sheet to represent weather information over the given
area. 
Note that obstacles and terrain information, being largely fixed, is already
implicitly incorporated in the learned model.}

\section{Large-Scale Simulations}
\label{sec:simulations}

In this section, we discuss our large-scale simulation results conducted over a large geographical area (we discuss an outdoor testbed evaluation in the next section).
We start with describing the underlying
propagation model used in our simulations. \mgtc{All our developed software is open-source~\cite{github}.}

\para{Longley-Rice Propagation Model and Setting.}
To evaluate our techniques over a realistic propagation model, we use
the well-known
Longley-Rice ~\cite{chamberlin82} Irregular Terrain With Obstruction Model (ITWOM), 
which is a complex model of wireless
propagation based on many parameters including locations, terrain
data, obstructions, soil conditions, etc.
We use SPLAT!~\cite{splat} to generate path-loss values; SPLAT! is an open-source software implementing the Longley-Rice propagation model.
When simulating the above propagation model, we consider an area of 
1km $\times$ 1km in our state and use the 600 MHz band to generate path losses
using SPLAT!.  
As the height of an entity is an important factor in determining the
path loss, we place the transmitters (PU or SU) at a height of 30m and the 
receivers (PURs and SSs) at 15m above the ground 
level.\footnote{\iam{At much lower heights, transient obstacles 
such as vehicles and temporary structures, would affect the path-loss 
model---which the SPLAT! software doesn't account for; thus, we choose
a higher altitude to simulate an \iar{accurate} setting.}}
For clarity of presentation and due to limited space, 
in many plots, we only show results for the
\pset; in these cases, the observed trend in \sset is similar.

\para{Performance Metrics (\perr, \pfp) and Algorithms.}
The main performance metric used to evaluate our technique is the average (absolute) difference in 
power allocated to the requesting SU with respect to an optimal algorithm;
here, the optimal algorithm has the knowledge of the exact path-loss values 
and is thus able to use the Eqn.~\ref{eq:maxpower} 
to compute the power to be allocated to the SU. 
We denote this measure by \perr. 
To evaluate the impact of false positives, we also consider the 
average false-positive
error metric \pfp which is computed as the aggregate error in the false-positive
samples over the false positives divided by the total number of samples. The \pfp metric
 informally measures the level of interference caused by the spectrum allocations.
We implement standard ML techniques, viz., a 3-layered neural network (\nn), 
support vector regression (\svr)~\cite{drucker1997support}, 
our CNN-based approaches \sname (\S\ref{sec:basic-cnn})
and \name (\S\ref{sec:deep-cnn}). Recall that \name is based on a pre-trained
deep CNN architecture \bluee{based on ResNet~\cite{resnet} architecture.}  
As detailed in \S\ref{sec:related}, the closest work relevant to our setting
is the one in~\cite{gupta19} which has available 
the PUs' parameter values as well as the SS readings, and uses 
interpolation techniques to estimate path-loss values between relevant 
entities; we denote this approach by \ip.\footnote{In our implementation of \ip algorithm, we fine-tuned 
its internal path-loss exponent parameter $\alpha$ to get the best performance; 
we ended up using a value of 3.3.} 
We note that the simple {\tt Listen-Before-Talk} approach resulted 
in a \perr of a several 
10s of dB in our simulations, and has not been shown in the plots for 
clarity. The large \perr
value for the very conservative {\tt Listen-Before-Talk} approach (see \S\ref{sub:related}) is due to the fact that in
a large majority of evaluation samples SU is not allocated {\em any} spectrum power as some of the PUs' power is always received in a large fraction of the area.

\para{Training Samples.}
As described in \S\ref{sec:deep_learning}, there are two types of training samples used
in training our models: (i) pre-training samples used to pre-train the \name; these samples
are based on the log-normal propagation model with a path exponent computed from a small 
number of path-loss samples; we computed an exponent of 3.3 from 200 path-loss samples. 
(ii) {\em training samples}, which are gathered 
in the field as described
below; these samples are used to train all the ML models, including the \name after pre-training. In addition, for our schemes, we also use synthetic samples (\S\ref{sub:cost}), 
which are derived from the training samples.
To create a training/evaluation sample, we place 10 to 20 PUs at random locations
in the given area and assign them a random power within the range of 0 to -30dBm. 
Each PU has a random number (5 to 10) of PURs distributed randomly within a distance of 
50m from the PU.
We vary the number of sensors from \bluee{49 to 625 (with 400 as the default)} 
sensors uniformly distributed in the field. 
For \nn and \svr models, we assume the maximum number of PUs to be 20, and use 
dummy parameter values if the number is less than 20. 

\tmagenta
\softpara{Validation and Evaluation Samples.}
In most of our plots, we vary the total number of training 
samples from 256 to about 4096 (with a default value of 2048), of which 
20\% are used for validation purposes (to tune the model's hyper-parameters) 
and the remaining are used to train the model.
In addition, we create and use 40,000 separate samples for evaluation purposes.
\tblack

\para{Training CNN models.}
\bluee{To pre-train the \name using the generated pre-training samples 
(we used 1m samples), 
we used SGD optimizer with a mini-batch size of 64. The learning rate starts from 0.01 and is divided by every 10 epochs. We use a weight decay of 0.00001 and a momentum of 0.9. 
Then, for fine-tuning the \name model using real (field) samples, 
we use \textit{Adam}~\cite{kingma2014adam} 
optimizer with a very low initial learning rate of 0.00001
while keeping the other parameter values the same.
Among all the hyper-parameters, we only optimize the regularization value.}

\subsection{Performance Results (Single SU)}
\label{sec:eval-single}

We now present our performance results for various algorithms and settings. We evaluate
various techniques for the case of a single SU request; we consider multiple SUs in the
next subsection. \bluee{\em Each data point in the plots is an average of over 40,000 evaluation samples.}

\begin{figure}%
    \centering
    \subfloat{{
        \includegraphics[width=.98\linewidth]{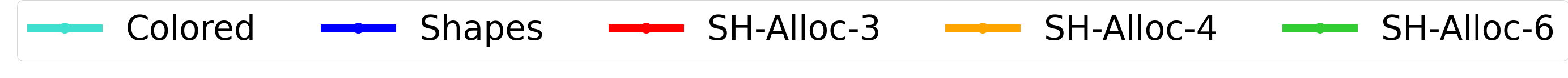} 
        }}
    \setcounter{subfigure}{0}
    \subfloat[\pset.]{{
    \includegraphics[width=.45\linewidth]{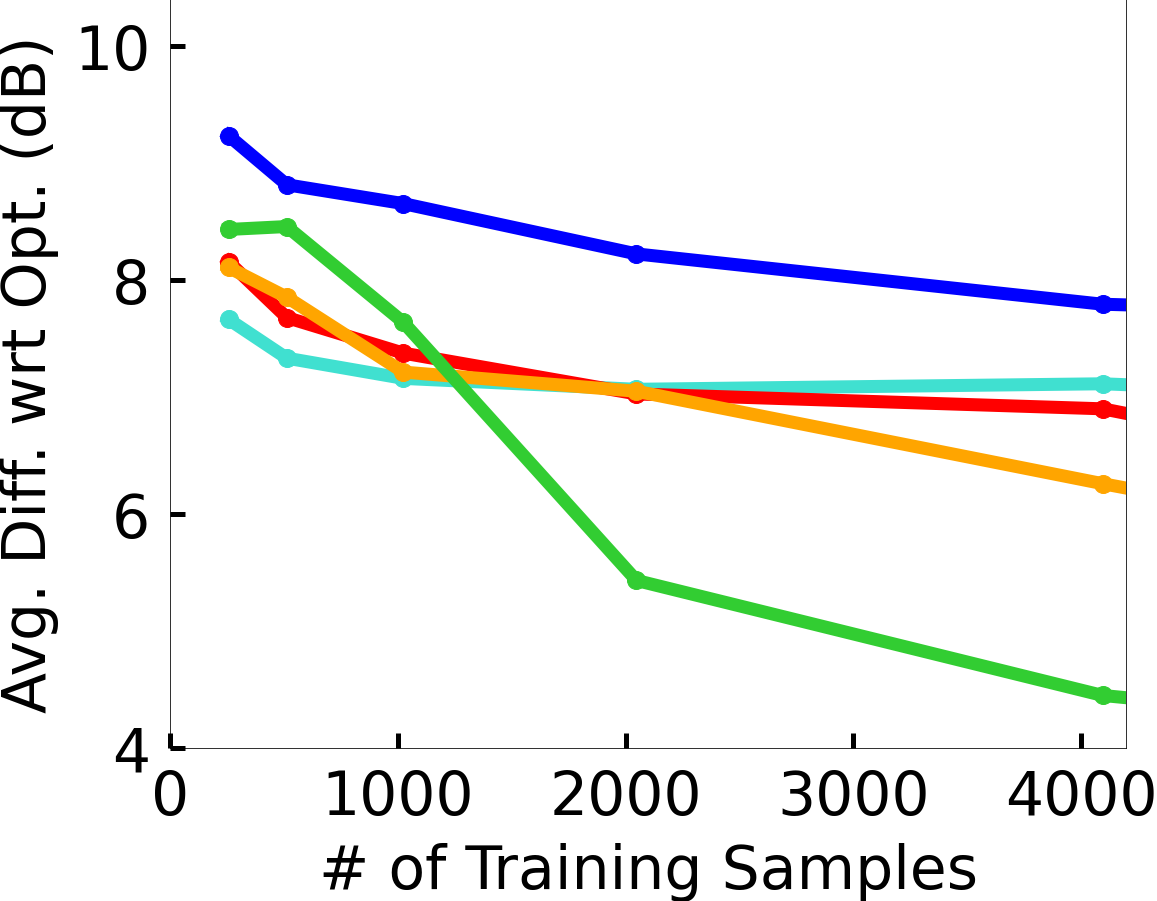} 
    }}%
    \qquad\hspace{-0.2in}
    \subfloat[\sset.]{{
    \includegraphics[width=.45\linewidth]{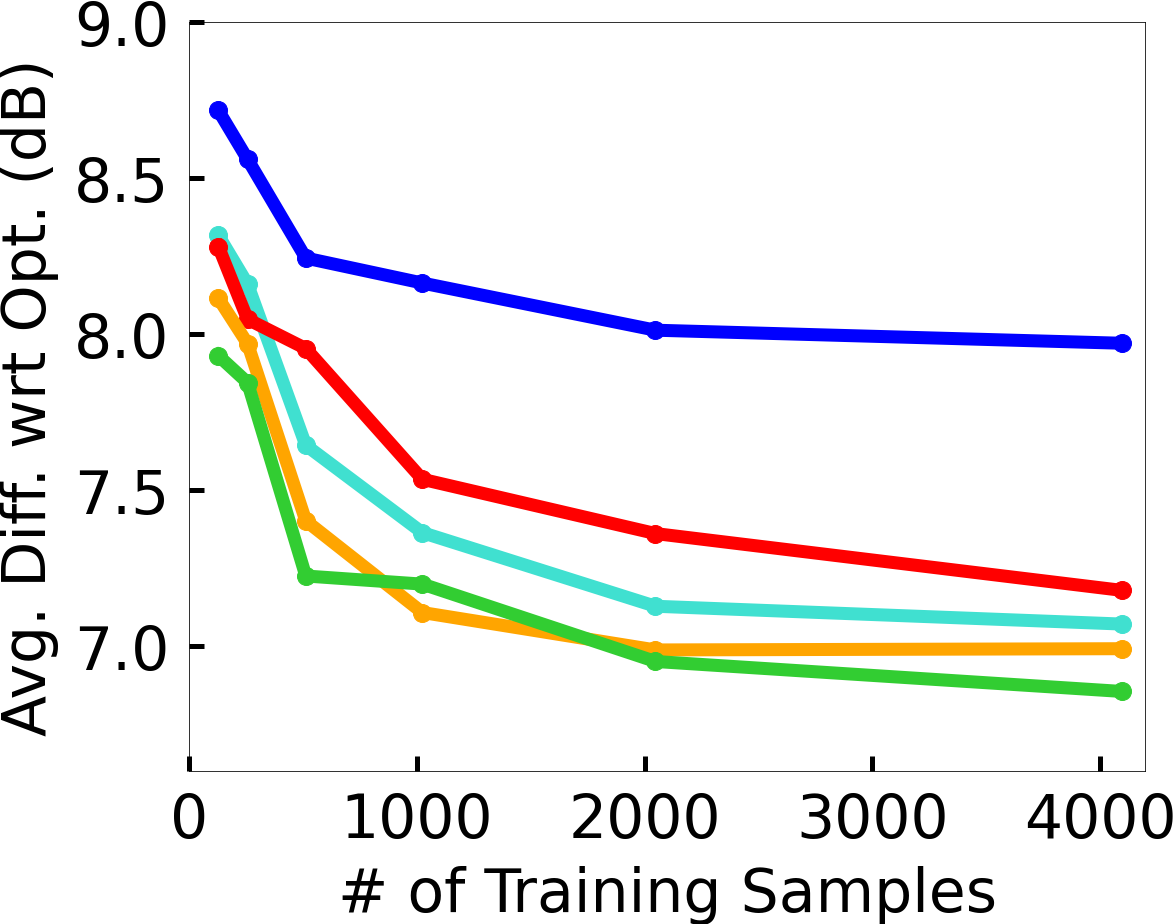} 
    }}%
    \caption{Performance comparison of various schemes for pre-processing the training samples into images.} 
    \label{fig:multiple_processing}%
\end{figure}
\para{Various Pre-Processing Schemes.}
We start with evaluating various pre-processing schemes for creating images from 
the training samples, as input to our basic CNN approach \sname. 
Here, we refer to the multi-sheets based approaches (i.e., from Fig.~\ref{fig:cnn_input}) as 
\sname-$N$, where $N$ is the number of sheets used for the PUs/SSs.
We also evaluate the {\tt Colored} and {\tt Shapes} schemes, where the 
\textit{Colored} scheme uses different colored-disks (red, green, and blue) for the 
three entities (PU, SS, and SU) and the {\tt Shapes} scheme uses a grey-scale image with different shapes for the entities (PU: circle, SU: square, SS: rectangle). 
We maintain the shape sizes to be small (to avoid intersections) and uniform for
each entity, but vary the intensity to represent the transmitted (received) power.
See Fig.~\ref{fig:multiple_processing} for a performance comparison of the above
schemes in both settings. We observe that the \sname-$N$ schemes outperform the other 
two schemes, with the increase in $N$ improving the performance as 
expected. 
Thus, we use 6 image sheets for PUs/SSs.

\begin{figure*}[t]%
    \begin{minipage}[t]{1\textwidth}
    \centering
    \subfloat{{\includegraphics[width=0.6\textwidth]{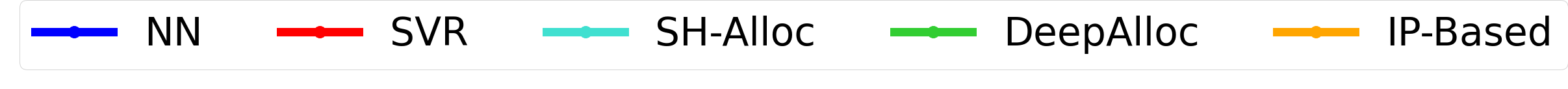} }}
    \end{minipage}
    \begin{minipage}[t]{1\textwidth}
    \centering
    \setcounter{subfigure}{0}
    \subfloat[\pset with increasing number of training samples and 10-20 PUs.]{{\includegraphics[width=.225\textwidth]{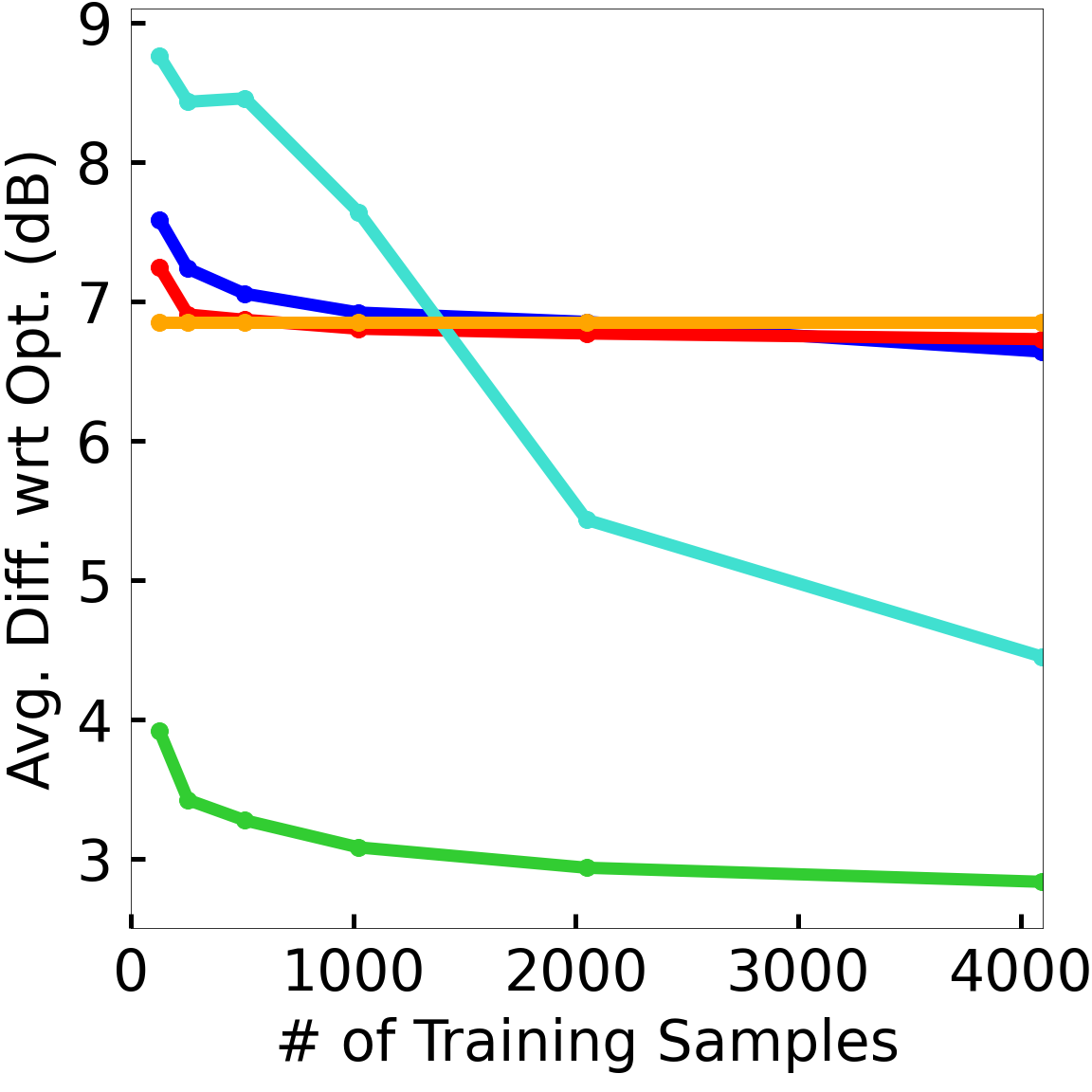} }}%
    \qquad\hspace{-0.16in}
    \subfloat[\pset with increasing number of PUs and 2k training samples.]{{\includegraphics[width=.225\textwidth]{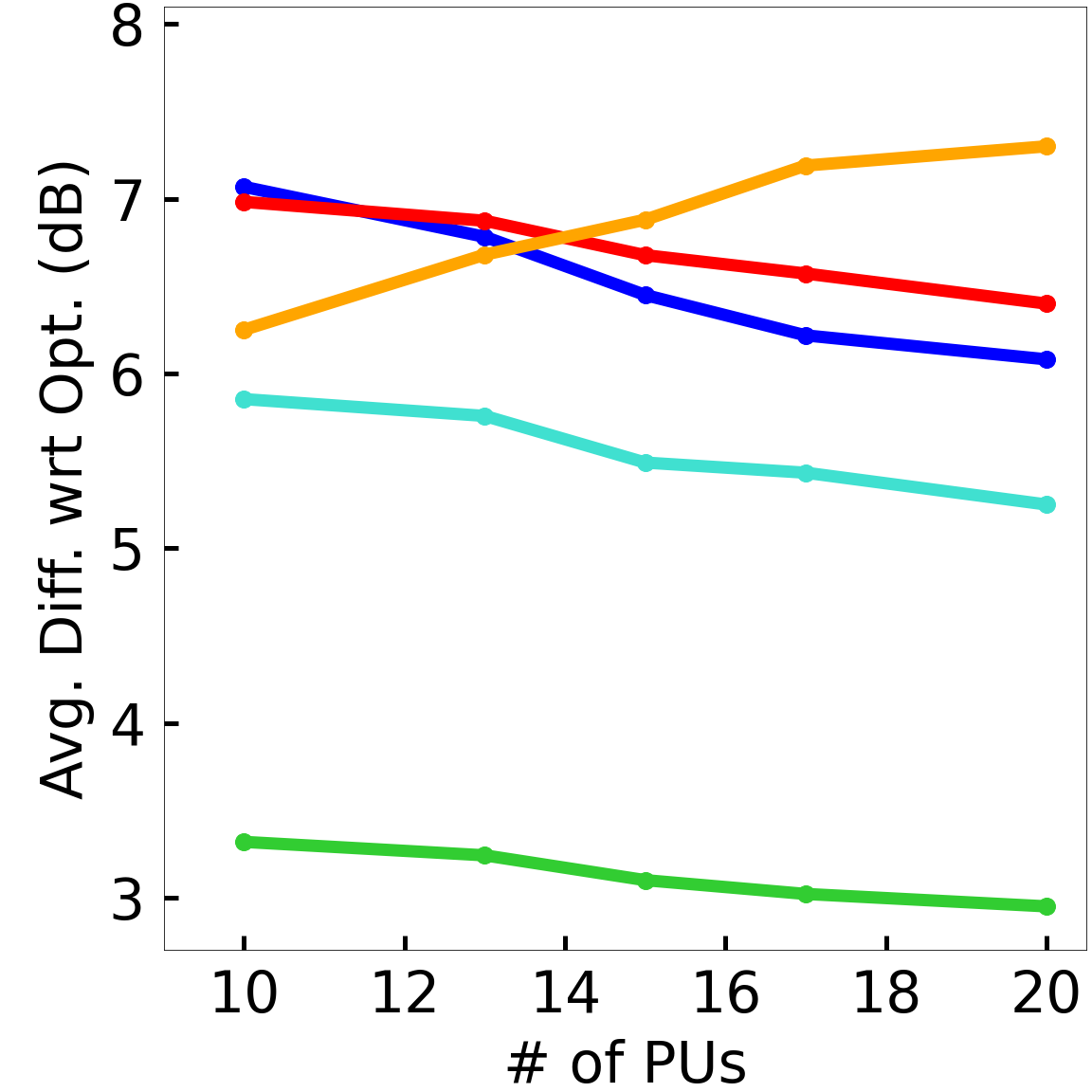} }}%
    \qquad\hspace{-0.16in}
    \subfloat[\sset with increasing number of training samples and 10-20 PUs.]{{\includegraphics[width=.225\textwidth]{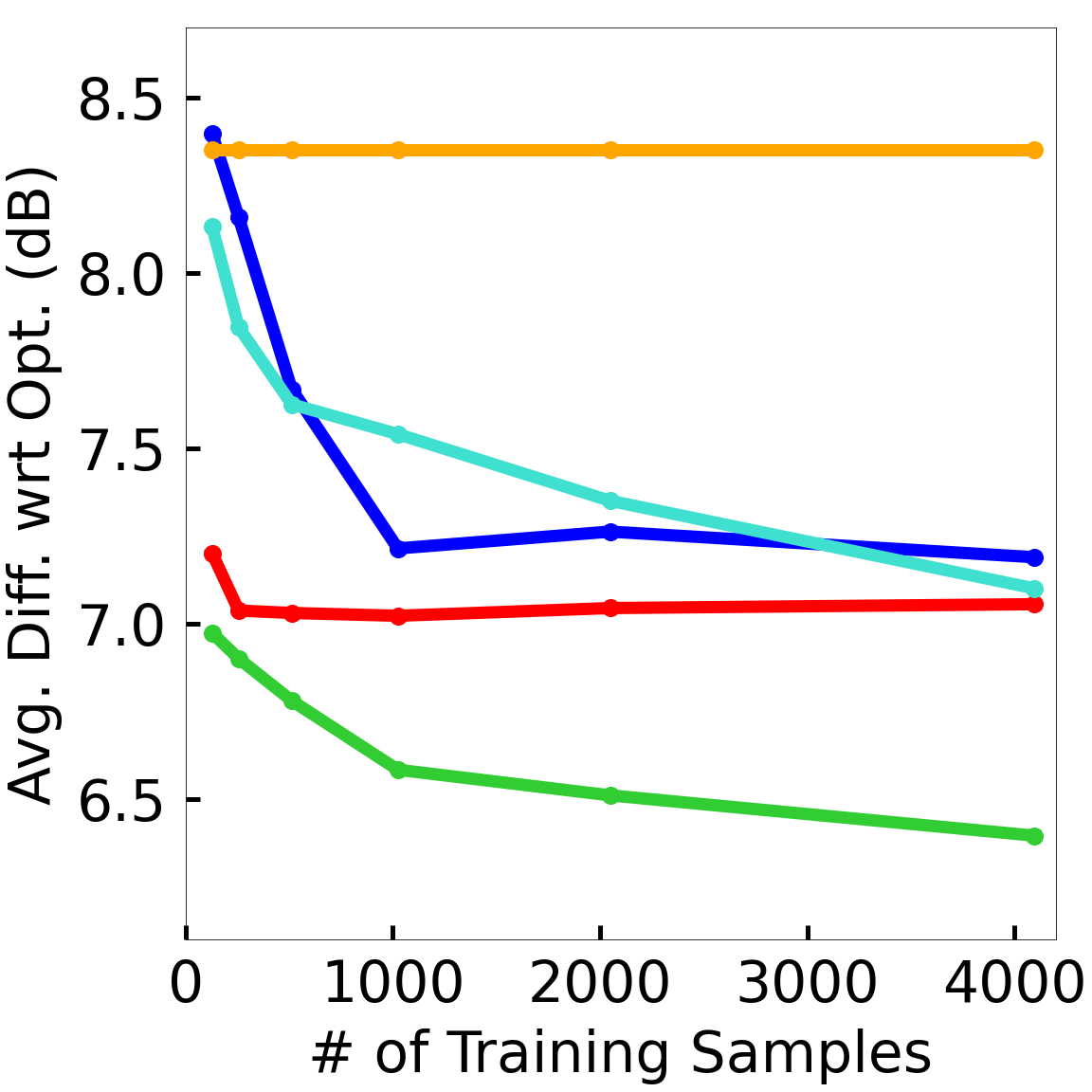} }}%
    \qquad\hspace{-0.16in}
    \subfloat[\sset with increasing number of SSs and 2k training samples..]{{\includegraphics[width=.225\textwidth]{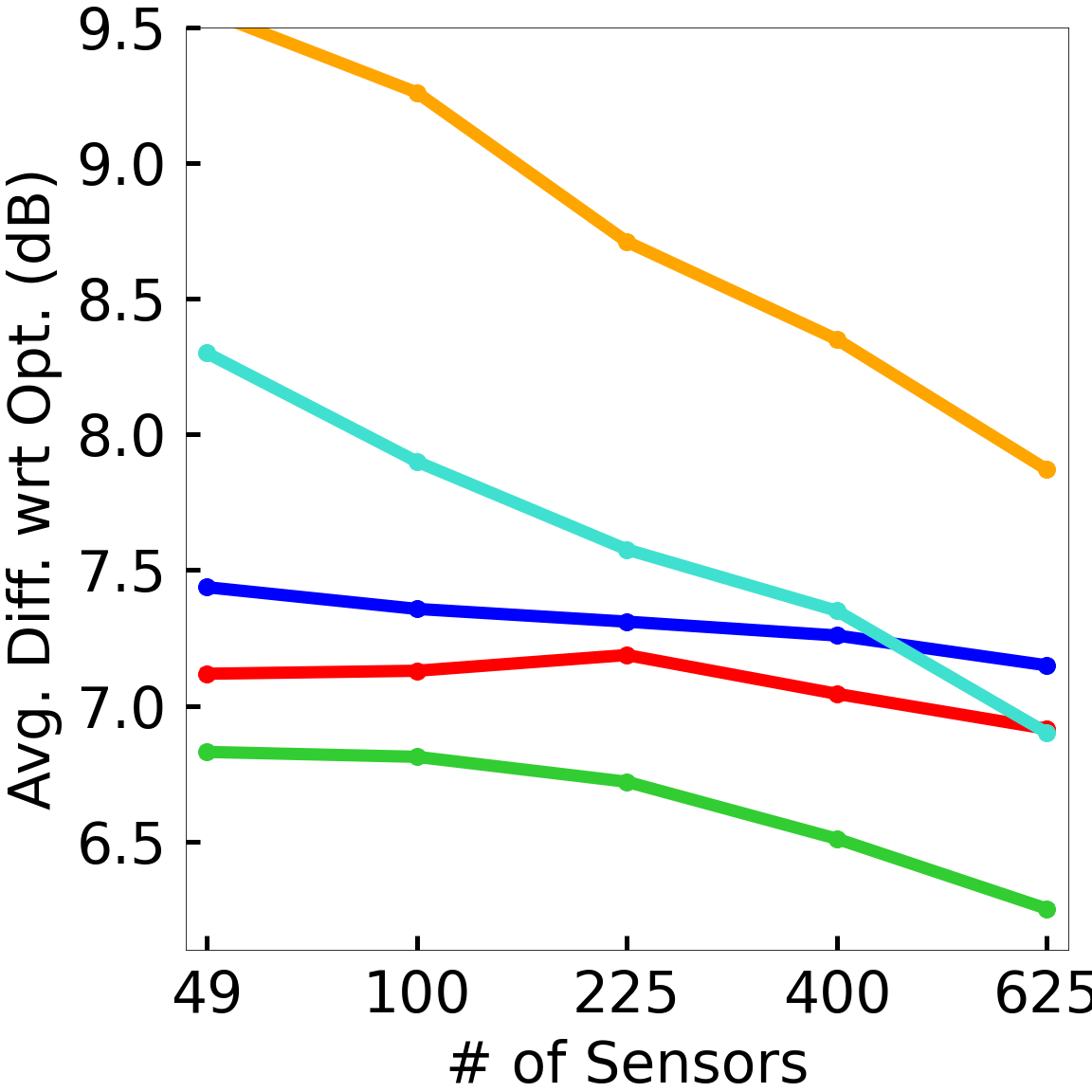} }}%
    \caption{Performance comparison of various algorithms for increasing number of training samples and PUs/SSs.}
    \label{fig:splat}
    \end{minipage}
\end{figure*}

\begin{figure*}[t]%
    \centering
    \subfloat[Minimizing \pfp.]{{\includegraphics[width=.225\linewidth]{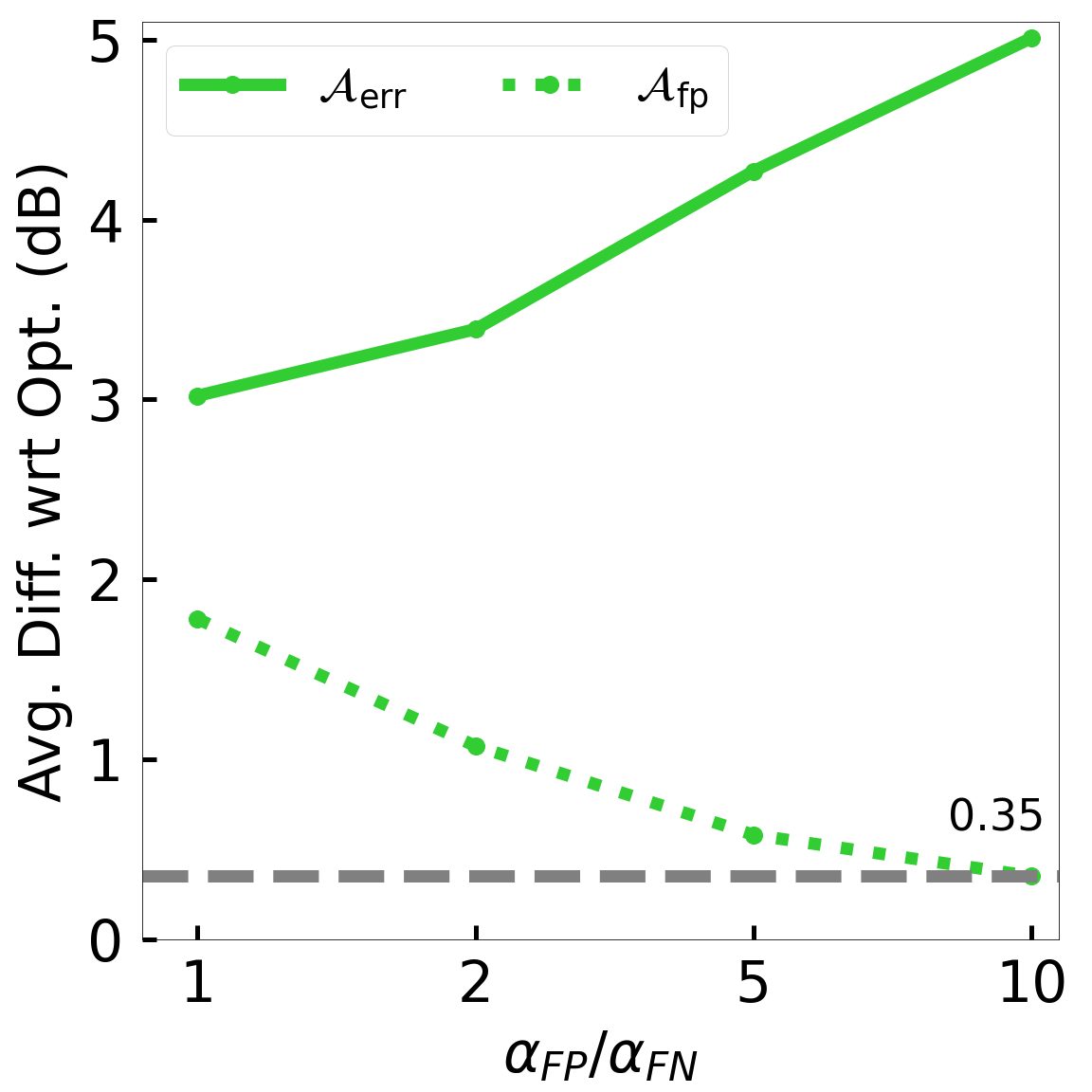} }}%
    \qquad\hspace{-0.2in}
    \subfloat[Handling Multi-Path Effects.]{{\includegraphics[width=.225\linewidth]{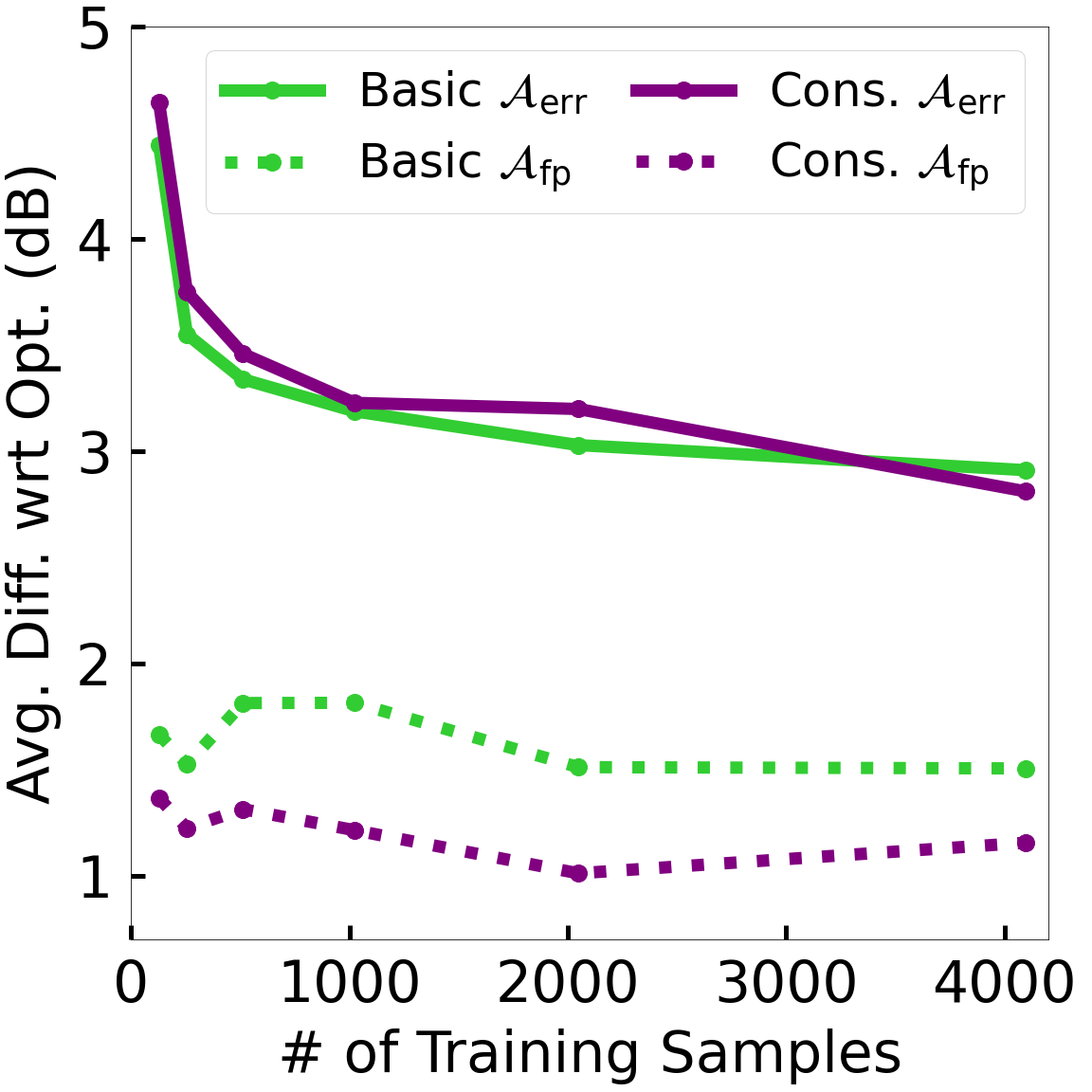} }}%
    \qquad\hspace{-0.2in}
    \subfloat[Synthetic samples in \pset.]{{\includegraphics[width=.225\linewidth]{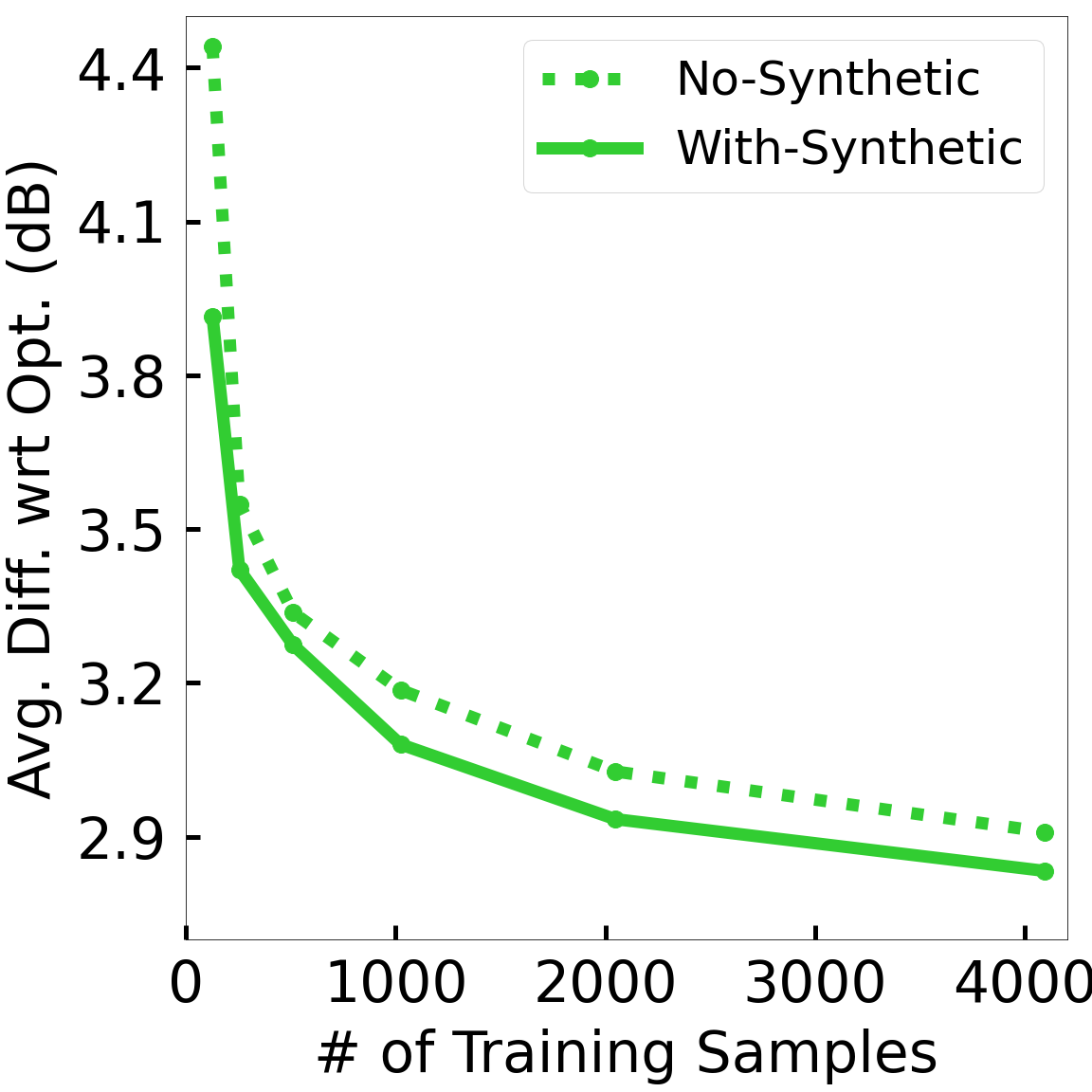} }}%
    \qquad\hspace{-0.2in}
    \subfloat[Synthetic samples in \sset.]{{\includegraphics[width=.225\linewidth]{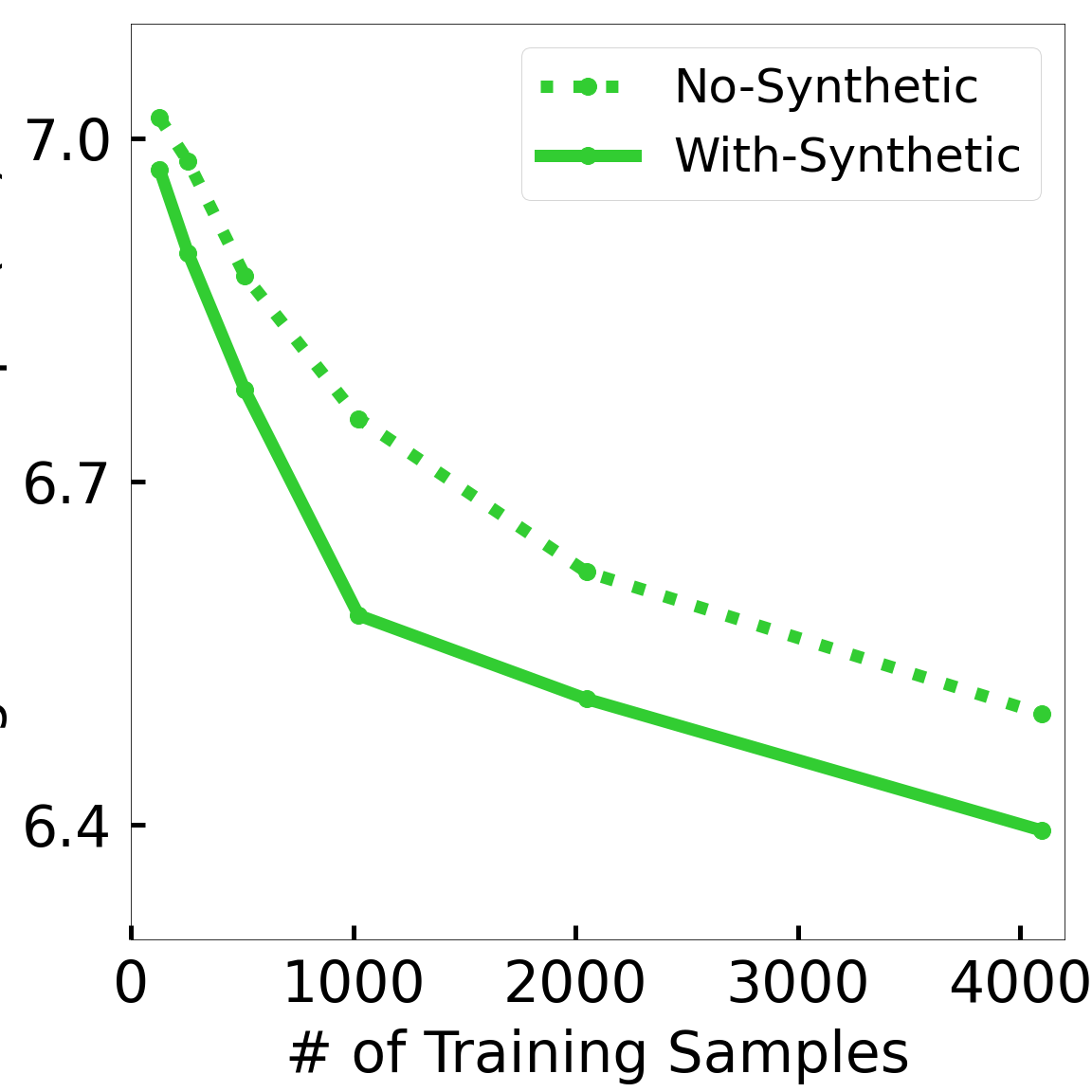} }}%
    \caption{(a) Minimizing false-positive error (\pfp), (b) Handing multi-path effect via learning a more conservative function, and (c)-(d) Using synthetic samples to improve performance. Here, (a)-(b) are in \pset; the \sset had similar results.}
    \label{fig:sub-techniques}%
\end{figure*}
\para{Varying Training Cost and Number of PUs/SSs.} 
We now evaluate the fundamental performance of various algorithms in terms of the key
\perr metric in our two settings, \pset and \sset, for varying number of (field-gathered)
training samples and \iar{the} number of PUs/SSs. See Fig.~\ref{fig:splat}.
First, our main \name approach outperforms all the other approaches for both settings.
More importantly, \name delivers great performance for the \pset with less than 3dB performance 
gap with the optimal solution. As expected, the performance of 
our approaches is better in \pset than that in \sset, 
since fundamentally the \pset has more direct information relevant to the 
spectrum allocation function.
Second and more specifically, we observe that both the CNN-based approaches 
outperform the \ip algorithm by a large margin in the \pset; in
the \sset, the \name approach easily outperforms the \ip algorithm, except for large number
of sensors wherein both techniques perform similarly. Recall that 
the \ip approach has the advantage of knowing {\em both} the PU parameters as well as SS 
readings, while the ML algorithms have knowledge of only one of these inputs---thus, it is surprising and commendable that \name is able to outperform the \ip approach.
The above observations suggest that \name is able to learn the spectrum allocation function effectively.
Lastly, we observe that the performance gap between the CNN-based approaches (\sname and \name) is
significant---which demonstrates the value of pre-training a deep CNN model.

\para{Evaluating Techniques for Minimizing False-Positive Errors, Handing Multi-Path Effect, and Synthetic Samples.}
We now evaluate our specialized techniques developed in \S\ref{sub:interference}-\ref{sub:cost} for
various aspects. 
See Fig.~\ref{fig:sub-techniques}.
Fig.~\ref{fig:sub-techniques}(a) evaluates the 
\S\ref{sub:multipath}'s
technique to minimize false positives. We plot \perr and
\pfp metrics for varying \afp/\afn ratio; we see that increase in \afp/\afn ratio is effective in minimizing the
false-positive error \pfp, at the cost of higher average \perr.
Next, in Fig.~\ref{fig:sub-techniques}(b), we evaluate the technique from
\S\ref{sub:multipath} to handle multi-path fading effects by learning
a conservative function via modified labels. Here, we plot the \perr and \pfp 
for the basic as well as the conservative technique. 
We observe that the conservative technique's false-positive error (\pfp) is indeed
reduced without much increase in the total error (\perr), compared to the basic 
approach. 
Fig.~\ref{fig:sub-techniques}(a)-(b) are in \pset; the \sset results were similar (not shown).
Finally, in Fig.~\ref{fig:sub-techniques}(c)-(d), we evaluate the benefit of 
using synthetic samples as described in \S\ref{sub:cost}. Here, we show \pset as well
as \sset as the algorithms for generating synthetic samples are very different. 
In \pset (Fig.~\ref{fig:sub-techniques}(c)), the performance gain due to synthetic samples
in positive but minimal---perhaps, because the performance was already near-optimal.
In \sset (Fig.~\ref{fig:sub-techniques}(d)), we see a significant performance
improvement due to synthetic samples---validating our use of interpolation-based synthetic 
samples. Here, we added 2500 new uniformly distributed sensors and interpolated their readings
using the four nearest original sensors. 

\begin{figure}%
    \centering
    \setcounter{subfigure}{0}
    \subfloat[Performance across regions.]{{
    \includegraphics[width=.45\linewidth]{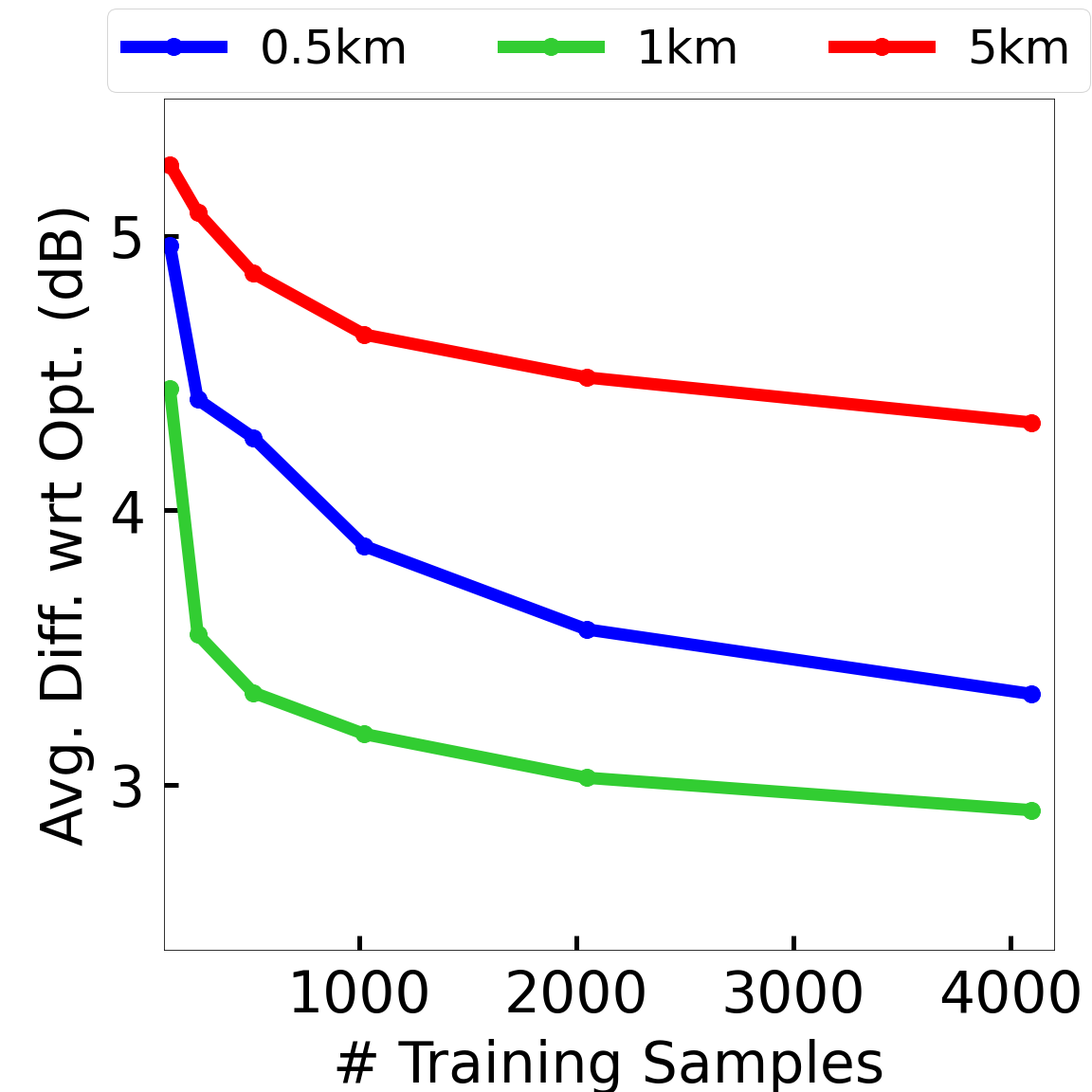} 
    }}%
    \qquad\hspace{-0.2in}
    \subfloat[Various pre-trained CNN models.]{{
    \includegraphics[width=.45\linewidth]{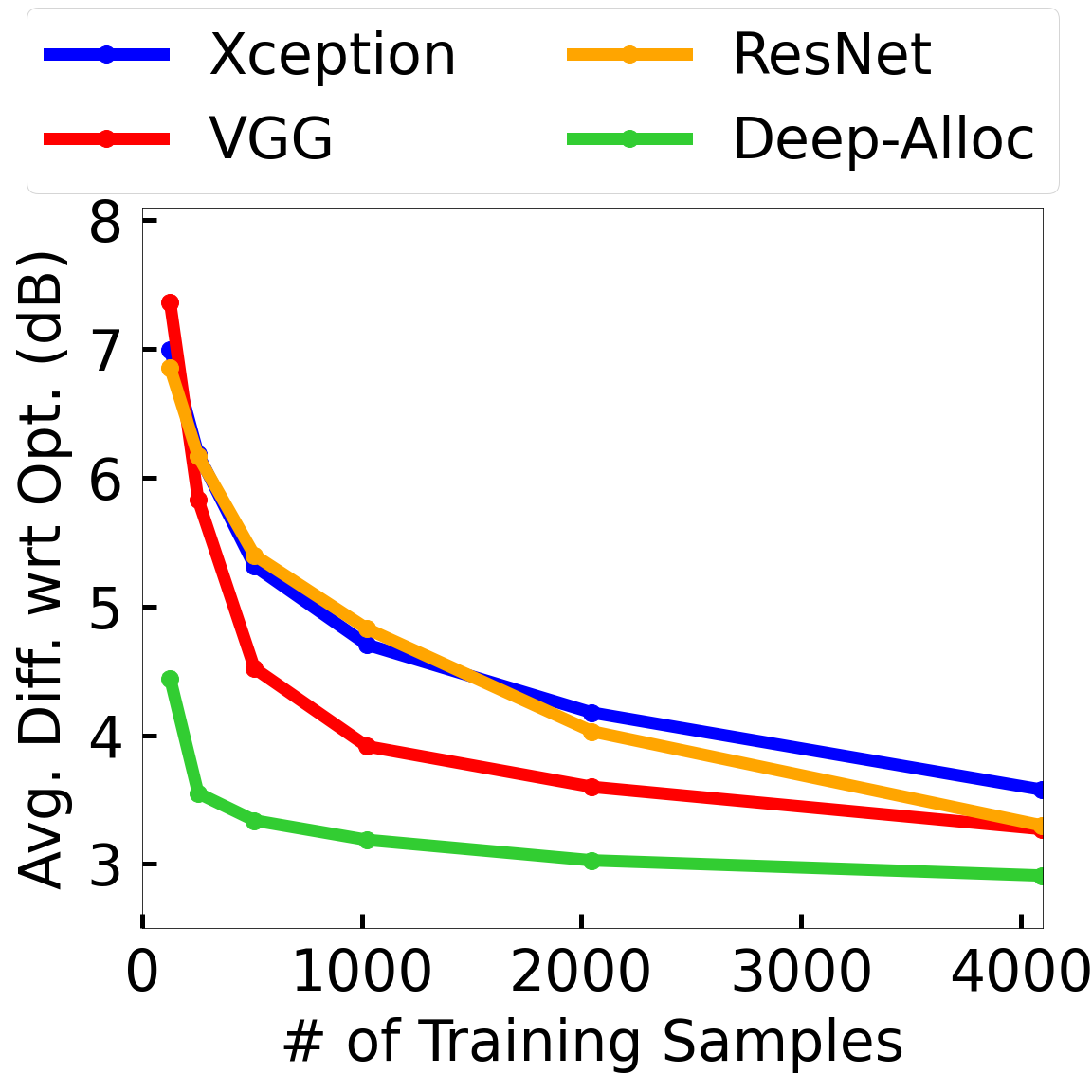} 
    }}%
    \caption{Performance comparison of (a) \name over different regions, and (b) various pre-trained models for increasing number of training samples.}
    \label{fig:multi_regions_pretrained}%
\end{figure}

\begin{figure*}[t]%
    \begin{minipage}[t]{0.25\textwidth}
    \setcounter{subfigure}{0}
    \subfloat[{\em Last} SU's Allocation.]{{\includegraphics[width=1\textwidth]{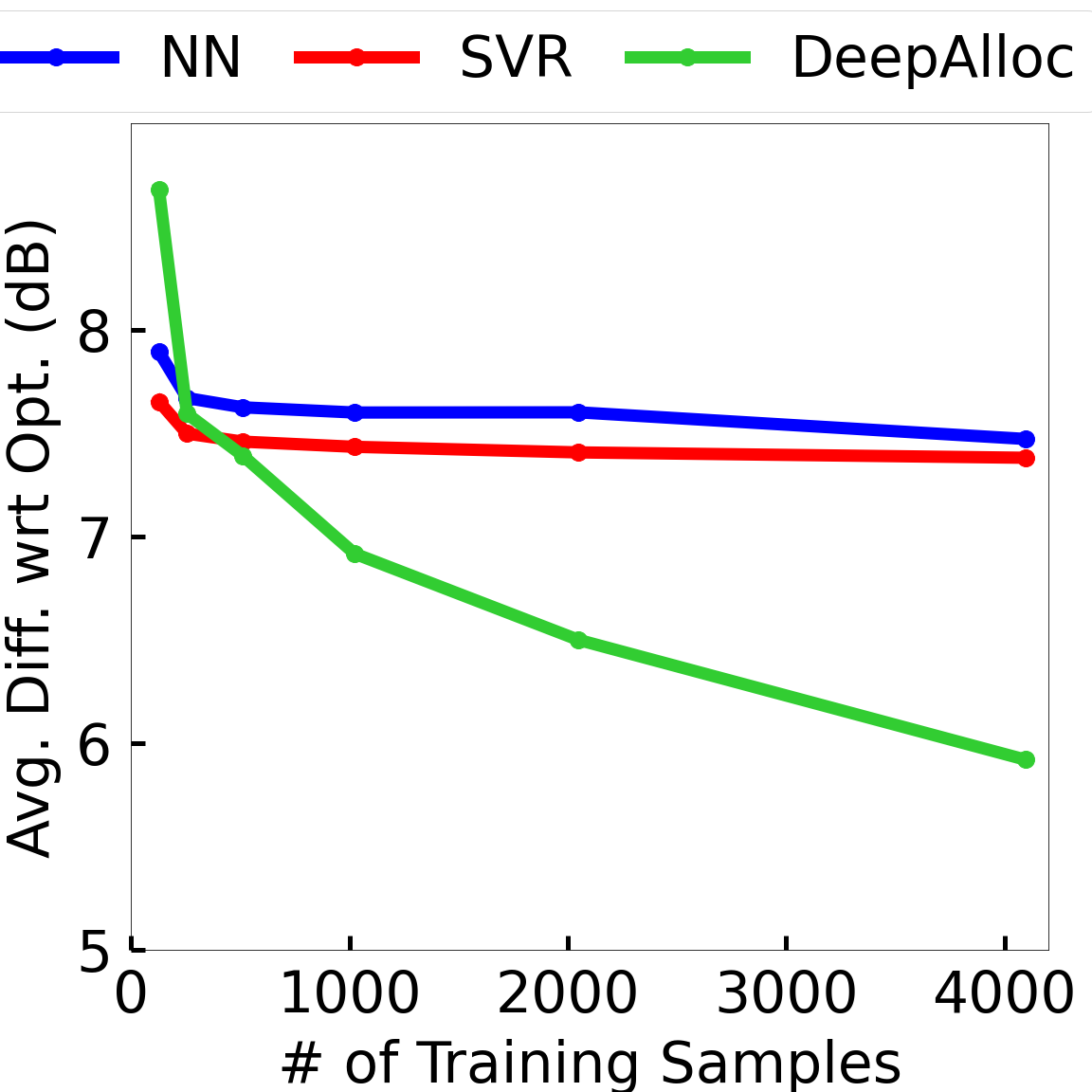} }}%
    \end{minipage}%
    \hfill
    \begin{minipage}[t]{0.75\textwidth}
        \vspace{-1.8in}
        \begin{minipage}[t]{1\textwidth}
            \centering
            \subfloat{{\includegraphics[width=0.6\textwidth]{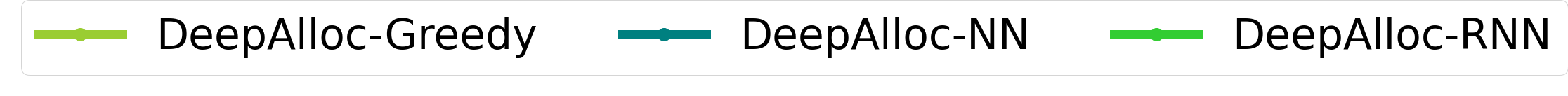} }}
        \end{minipage}
        \begin{minipage}[t]{1\textwidth}
            \setcounter{subfigure}{1}
            \subfloat[Average difference of aggregated SUs power ]{{\includegraphics[width=.30\textwidth]{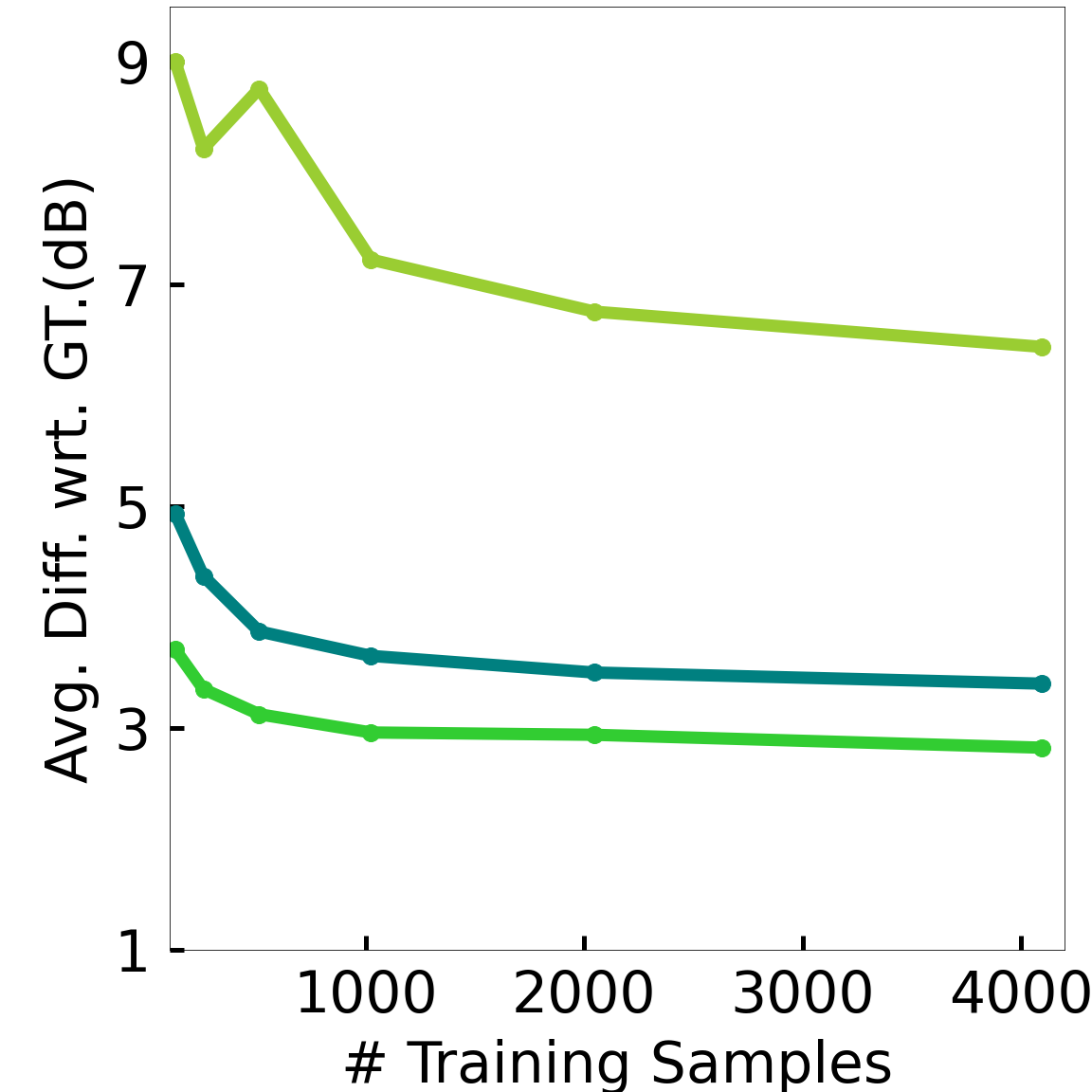} }}%
            \qquad\hspace{-0.2in}
            \subfloat[Fairness across multiple SUs.]{{\includegraphics[width=.30\textwidth]{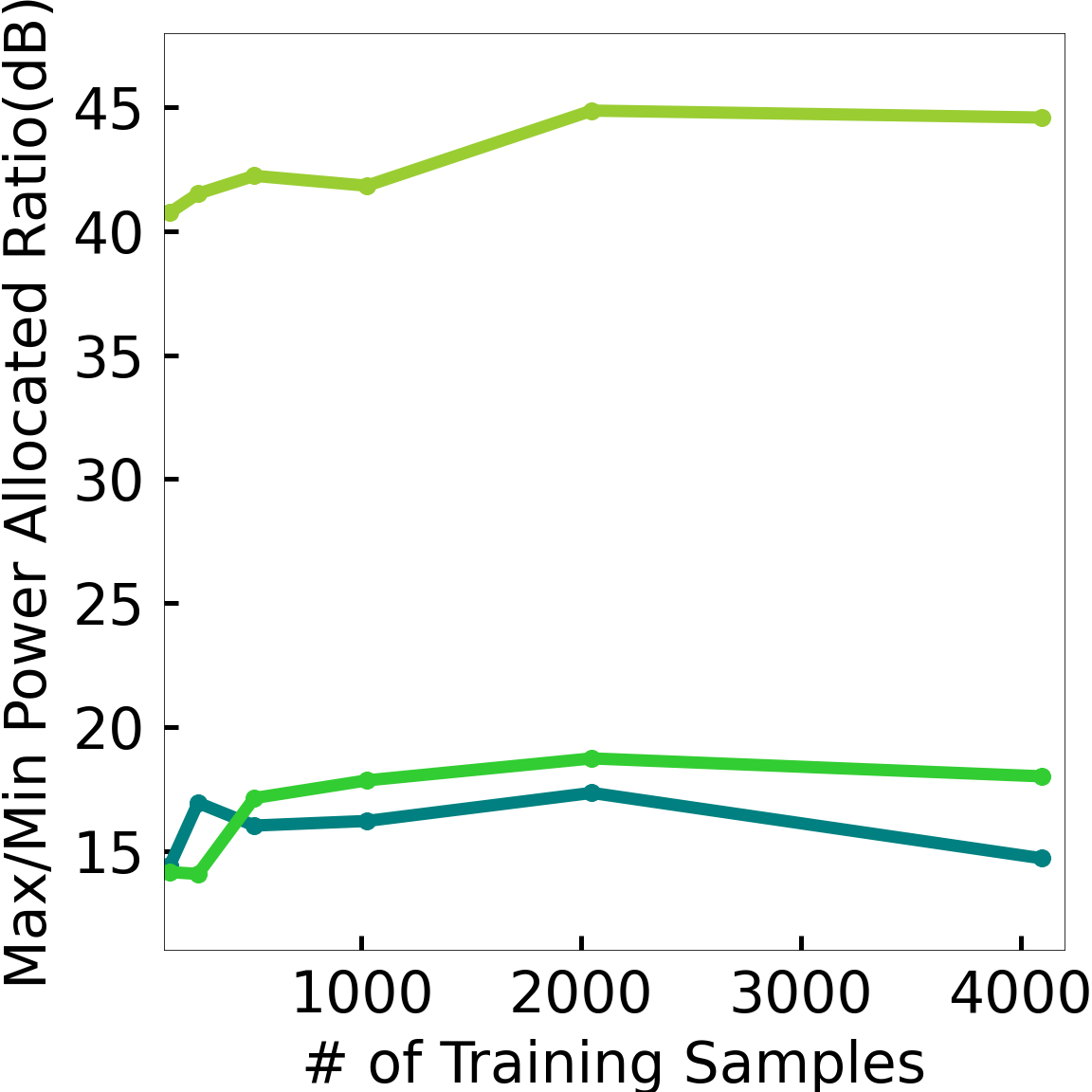} }}%
            \qquad\hspace{-0.2in}
            \subfloat[Average data rate by multiple SUs.]{{\includegraphics[width=.30\textwidth]{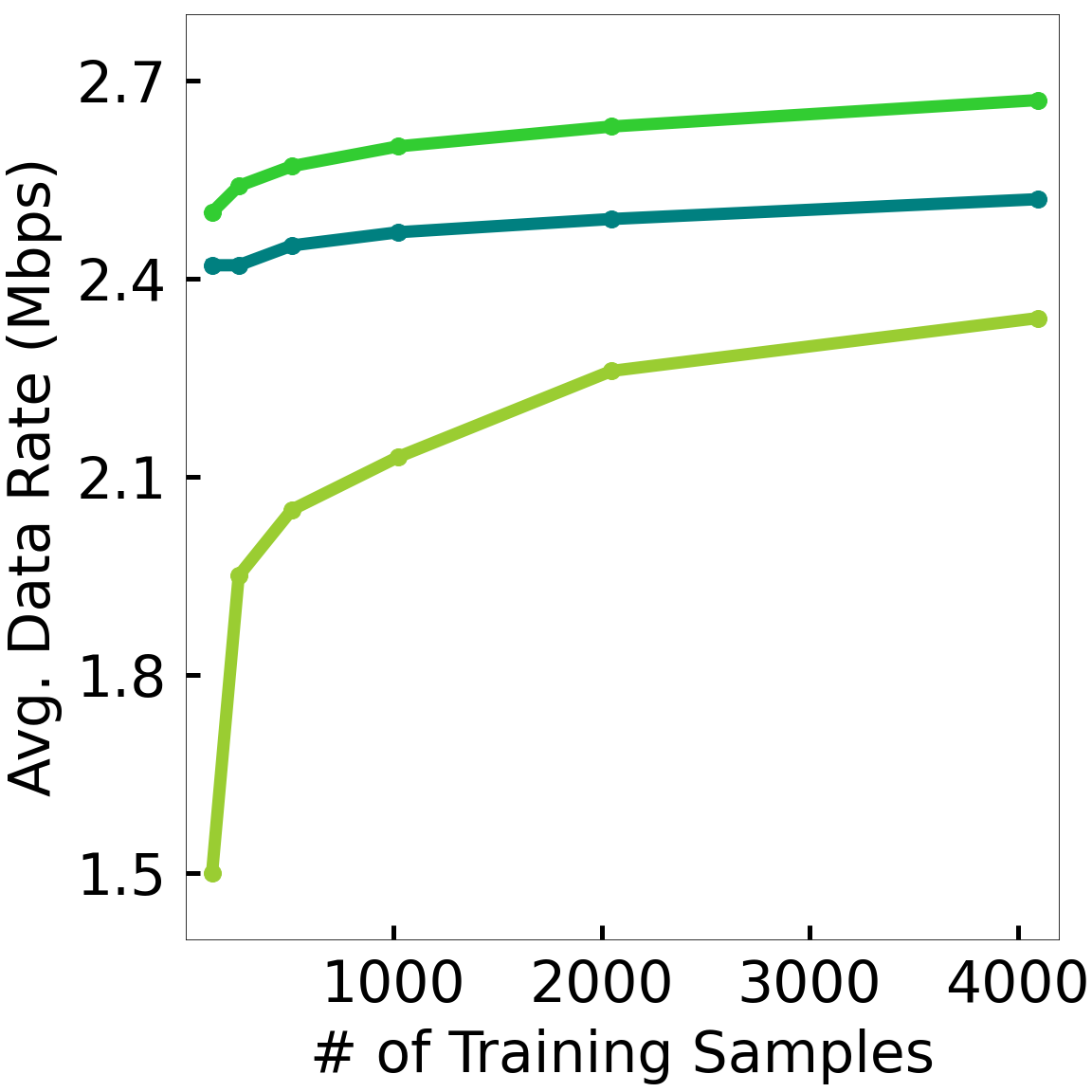} }}%
        \end{minipage}
    \end{minipage}
    \caption{Multiple SUs. 
    (a) Power allocated to subsequent SUs, in presence of other active SUs. 
    (b)-(d) Average different wrt ground truth (GT), Fairness (max/min ratio), and Total data rate for multiple SUs by various multi-SU algorithms.}
    \label{fig:fairness}
    
\end{figure*}

\para{Multiple Regions.}
We also evaluated our \name scheme over other regions with different sizes and terrain characteristics. See Fig.~\ref{fig:multi_regions_pretrained}(a), which plots performance
of \name over three different regions---$500m\times500m$ (our university campus), $1km\times1km$ (airport landing area, the default region in all other plots), and $5km\times5km$ (an urban area). 
We plot results for \pset wherein the number of PUs is between 10 and 20. We see that {\tt \name} has an \perr of 3-4.5 dB across all regions, and it performs the best over the airport landing area likely due to the lack of buildings and 
thus more uniform 
path-loss characteristics.

\para{Various Pre-Trained Models.}
Finally, we present evaluation results for various pre-trained deep models. In particular, in addition
to our \name architecture, we also used some well-known Image Classification models such 
as VGG~\cite{vgg}, ResNet~\cite{resnet}, and Xception~\cite{xception} which are all pre-trained with
over 1 million images involving our daily-life objects; these pre-trained models 
were further trained with \iab{a} varying number of (field-gathered) training samples.\footnote{For these models, we use {\tt Colored} scheme of creating images
from training images, as they use colored images (rather than sheets) during pre-training.}  
See Fig.~\ref{fig:multi_regions_pretrained}(b), which plots the \perr metric in \pset for the above models compared with our \name. 
We see that our approach of pre-training using log-normal model-based images in \name yields a notable performance improvement; the performance gap is particularly 
significant for a lower number of training samples. 

\tmagenta

\para{Computation Times and Complexity.}
The inference time complexity of all our ML approaches is linear in the size of the input, and thus, the inference time in practice is minimal (a fraction of a second). The training time complexity of most ML models depends on the training samples and the resulting convergence, and is thus, uncertain. The actual training times incurred from our set of 
training samples, on a 4GHz 8-core machine with a GeForce RTX 3080 GPU, were as follows: 
a few minutes for \nn as well as \svr approaches, and a few hours for \sname. To train the \name model, it took 5-7 days of computation time to pre-train the model using 1M images and a few hours to train/fine-tune the model after pre-training using SA training samples.
\tblack


\subsection{Multiple SUs/Channels}
\label{sec:evals-multi}


\para{Subsequent SU.}
We start with comparing 
the spectrum power allocated to a {\em single} new SU with previous SUs being active.
Each sample has a random number (between 1 and 10) of active SUs.
For evaluation purposes, we assume that the active SUs are assigned the optimal power (as per
Eqn.~\ref{eq:maxpower}, with full knowledge of PU information and path-loss values) and they 
transmit using this optimal power. 
Then, the models predict the allocation power for the last requesting SU. 
See Fig.~\ref{fig:fairness}(a), which shows the performance of various ML algorithms 
in assigning the power to the last SU. 
We see that \name easily outperforms the other ML approaches by a large margin.
Also, not shown for clarity, but \ip had a poor \perr of 21.6dB and \sname had a \perr's of 7.5dB.

\para{Concurrent SUs: Singular vs.\ Simultaneous Allocation.}
We now compare our approaches for handling concurrent SUs (\S\ref{sec:multiple_sus}), 
viz., \nameg, \namenn, and \namernn; recall that the latter two approaches handle SUs
simultaneously. We compare these algorithms in terms of (i) Average absolute-difference
wrt to the ground truth (i.e., the output of the \bina heuristic); 
(ii) Fairness, in terms of the ratio of maximum
to minimum power allocated to an SU; a lower value suggests a more fair allocation.
(iii) Overall spectrum utilization, in terms of the total data rate achievable which is estimated as follows. 
For each SU, we assign a receiver in a random location around the SU (within 100m), 
calculate the total interference from PUs and other SUs at the receiver, 
and estimate the achievable data rate using Shannon's capacity 
law.
See Fig.~\ref{fig:fairness}(b)-(d).
We observe that as expected the \namenn and \namernn approaches that allocate powers
simultaneously outperform \nameg significantly in all metrics, with \namernn slightly
outperforming the \namenn approach too (except in the fairness metric) 
which is not surprising as \namernn has much more input information. 



\begin{figure}%
    \centering
    \subfloat{{
    \includegraphics[width=0.7\linewidth]{figures/channels_legend.png} 
    }}%
    \setcounter{subfigure}{0}
    \subfloat[Total allocated power.]{{
    \includegraphics[width=.45\linewidth]{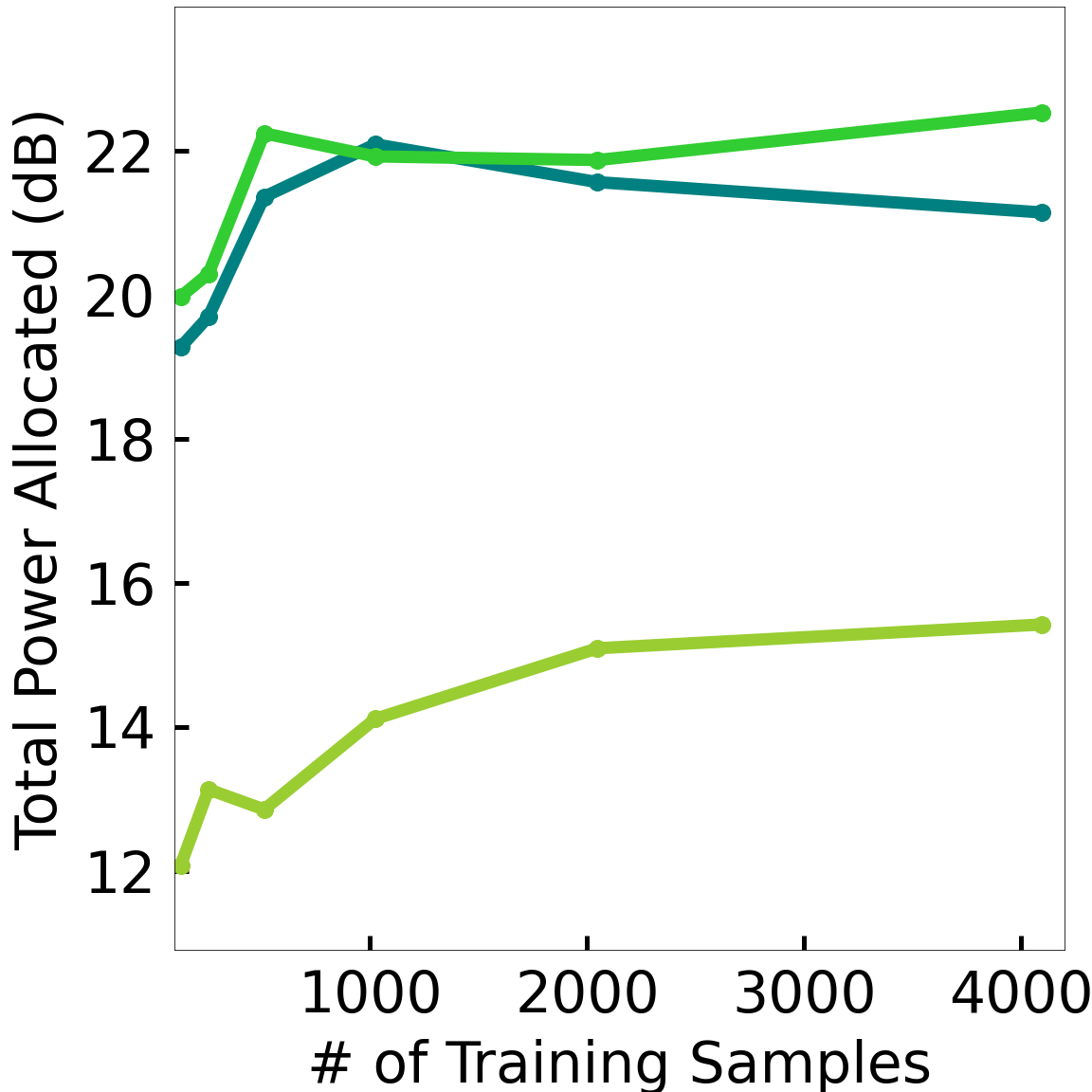} 
    }}%
    \qquad\hspace{-0.2in}
    \subfloat[Fairness metric.]{{
    \includegraphics[width=.45\linewidth]{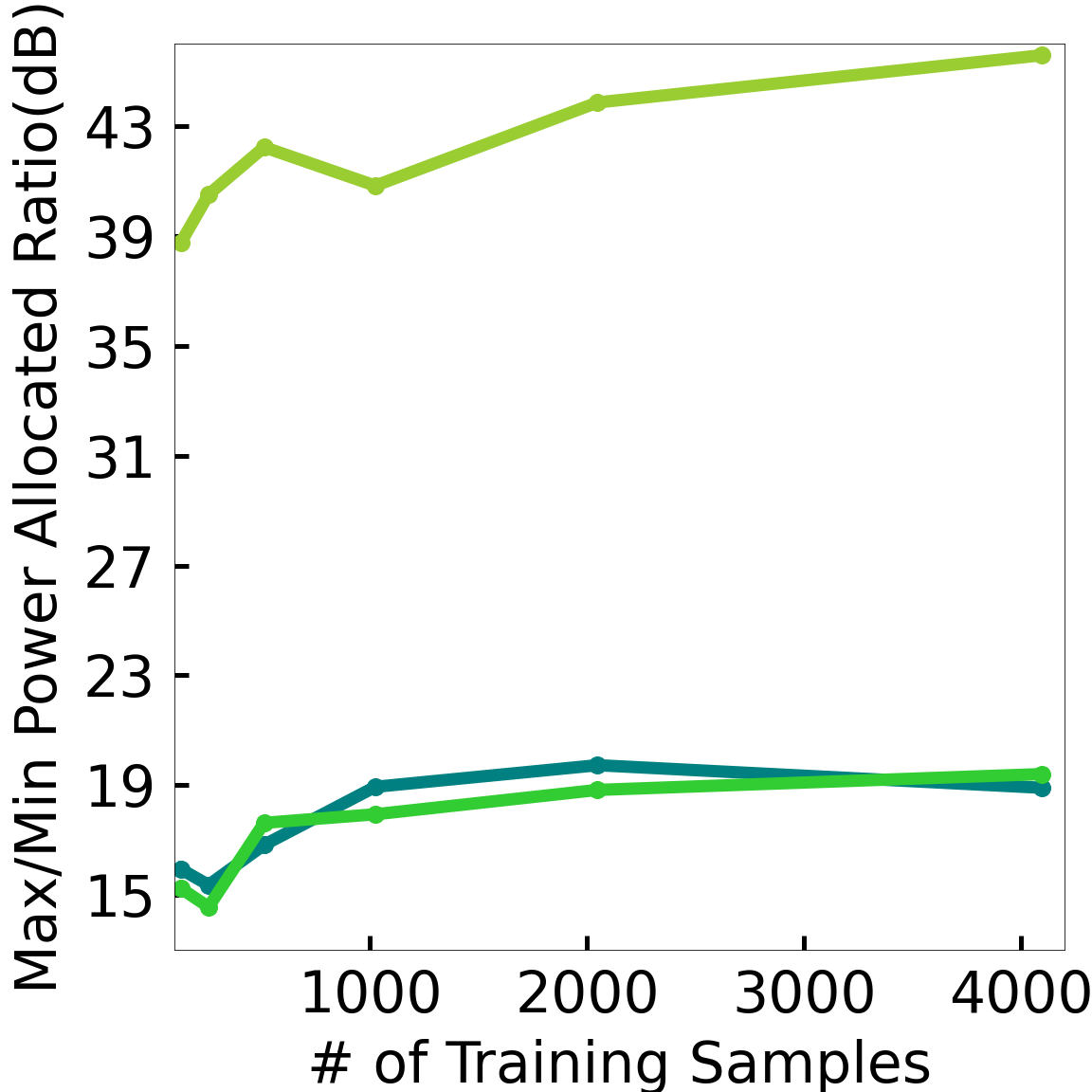} 
    }}%
    \caption{(a) Total allocated power, and (b) Fairness metric, in a four-channel setting with multiple SUs.}
    \label{fig:fairness_multi}%
\end{figure}
\para{Multiple Channels.}
In Fig.~\ref{fig:fairness_multi}, we compare our multi-SU approaches (tailored to multiple channels as described in \S\ref{sec:multiple_sus}) in the 4-channel \pset 
with PUs transmitting over all the channels at all times. We allocate powers 
to 30 SUs. We observe that
\namernn still achieves the best performance in terms of total power allocation as well
as fairness, with \namernn performing closely. With respect to the single channel, the total power allocated is almost four times (in Watts).

\begin{figure}
    \centering
    \subfloat[Devices]{
    \includegraphics[width=0.252\textwidth]{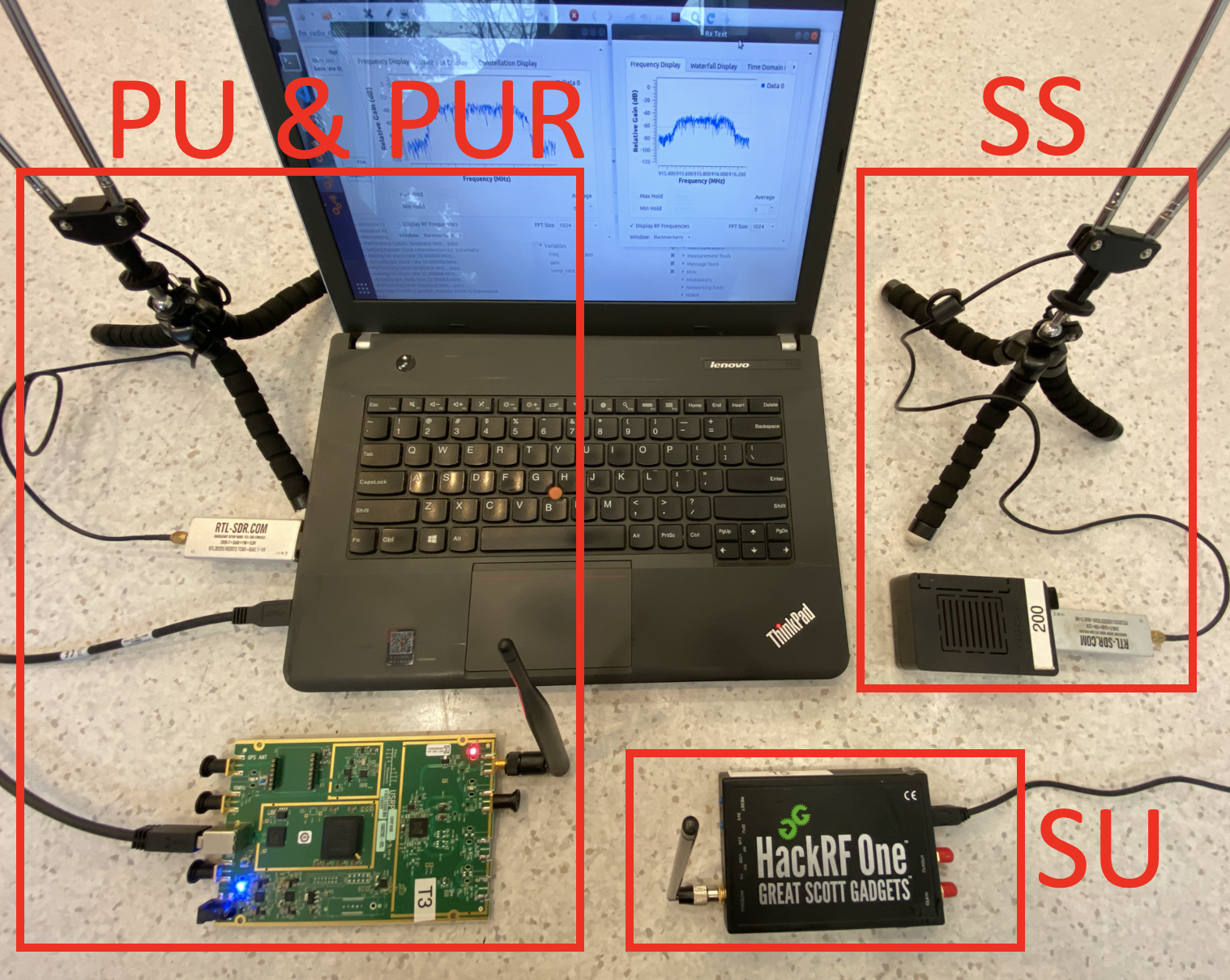} 
    }
    \subfloat[Testbed Area]{
    \includegraphics[width=0.216\textwidth]{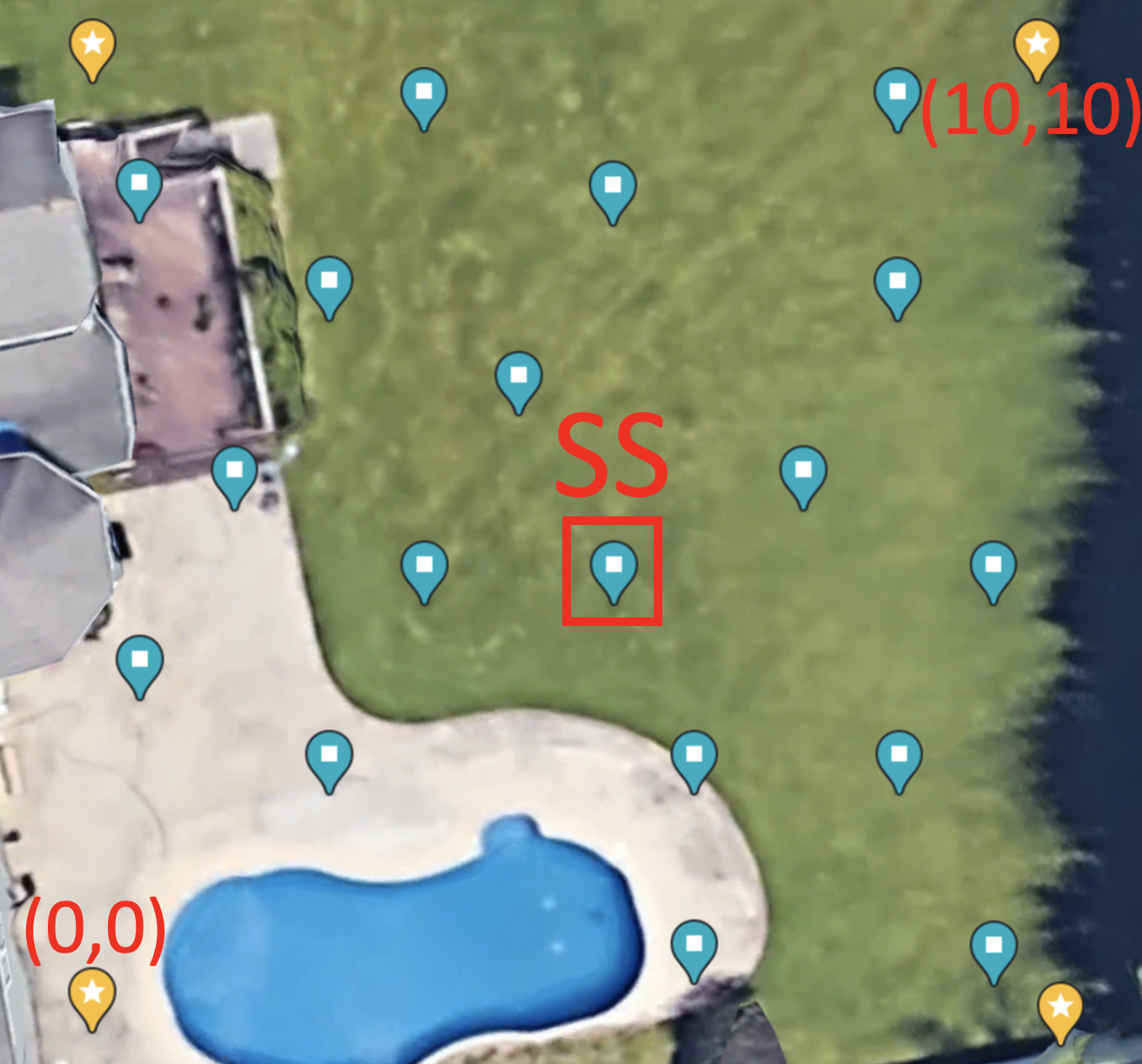} 
    }
    \caption{Outdoor testbed. (a) PU, PUR, SU, SS devices. (b) Testbed area (house backyard). Blue stars are the 17 sensors.}
    \label{fig:testbed}
\end{figure}

\begin{figure}
    \centering
    \includegraphics[width=0.49\textwidth]{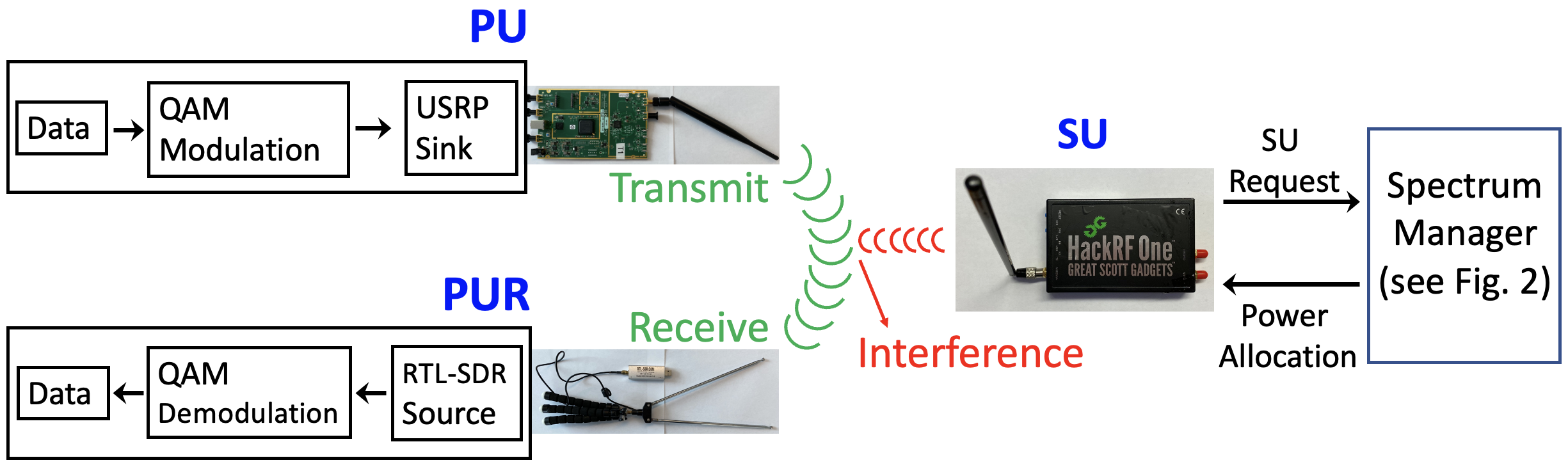} 
    \caption{Testbed system overview, including the GNU-Radio based PU-to-PUR communication system .}
    \label{fig:testbed_system}
\end{figure}

\begin{figure}%
    \centering
    \hspace{0.2cm}
    \subfloat{{
    \includegraphics[width=.9\linewidth]{figures/legends.png} 
    }}%
    \setcounter{subfigure}{0}
    \subfloat[\pset with 2-4 PUs.]{{
    \includegraphics[width=.45\linewidth]{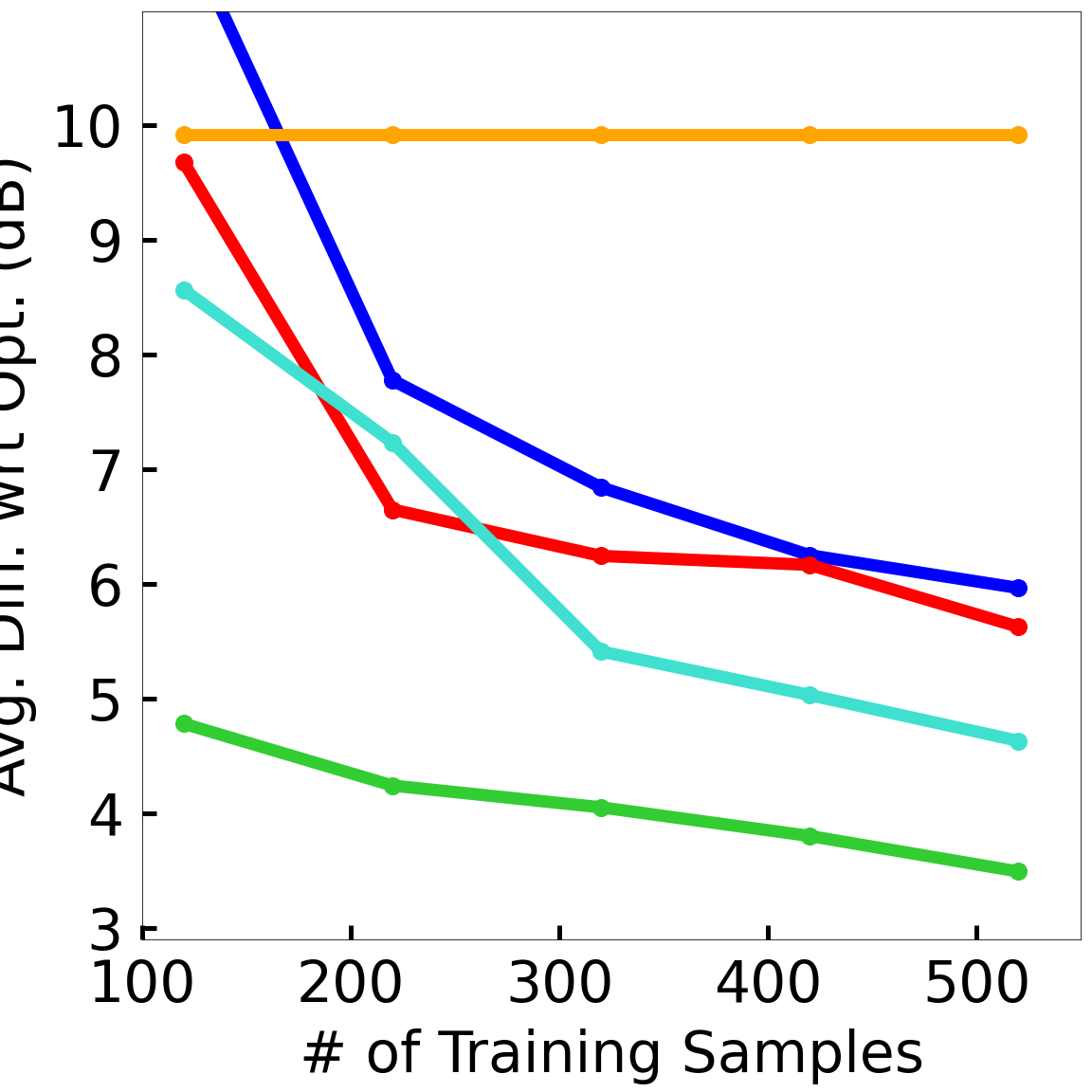} 
    }}%
    \qquad\hspace{-0.2in}
    \subfloat[\sset with 2-4 PUs, 17 SSs.]{{
    \includegraphics[width=.45\linewidth]{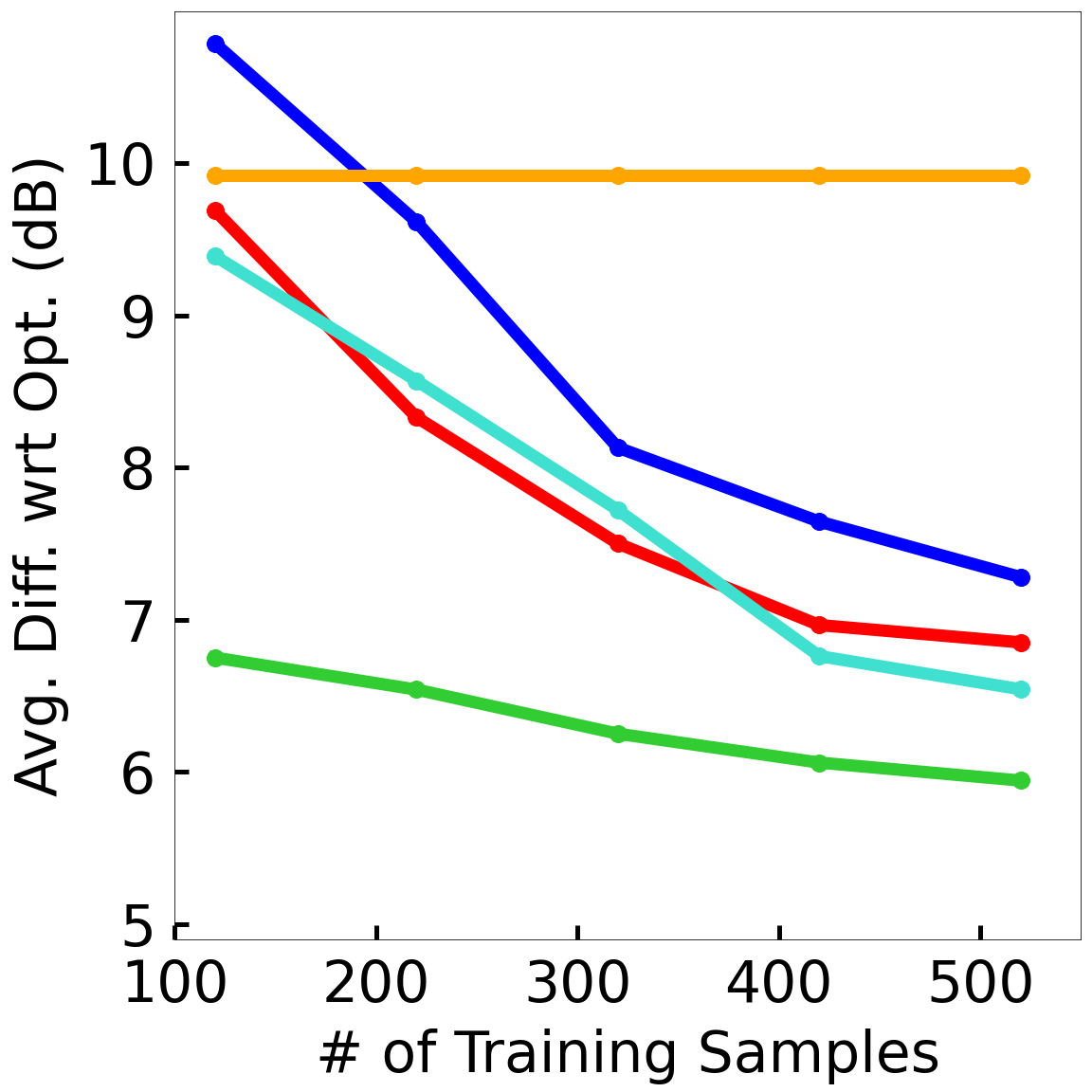} 
    }}%
    \caption{Testbed performance for increasing training set.}
    \label{fig:testbed_training_size}%
\end{figure}

\begin{figure*}[t]%
    \centering
    \begin{minipage}[t]{1\textwidth}
    \subfloat[Minimizing false positive error in \pset.]{{\includegraphics[width=.225\textwidth]{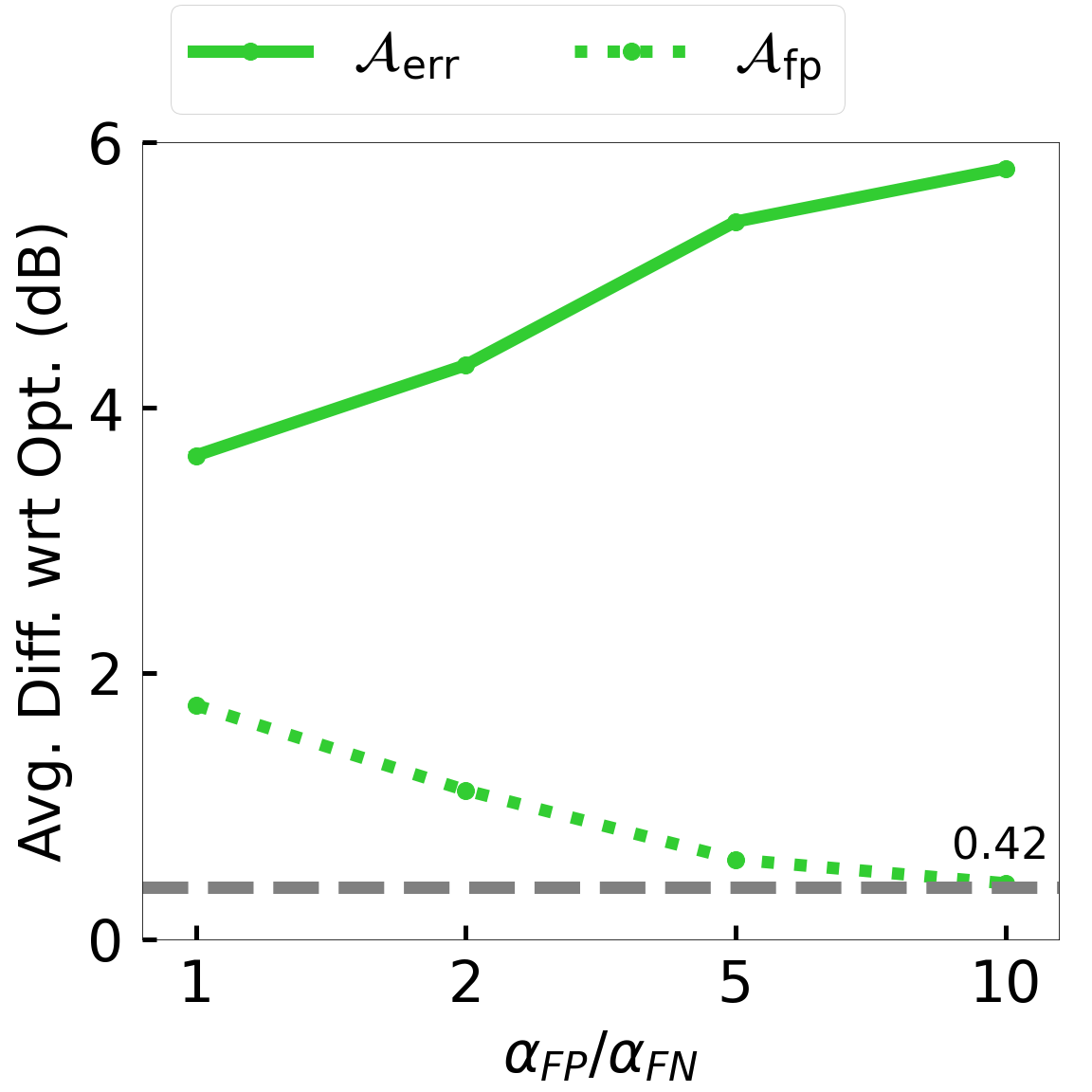} }}%
    \qquad\hspace{-0.16in}
    \subfloat[Handing multi-path effect in \pset.]{{\includegraphics[width=.225\textwidth]{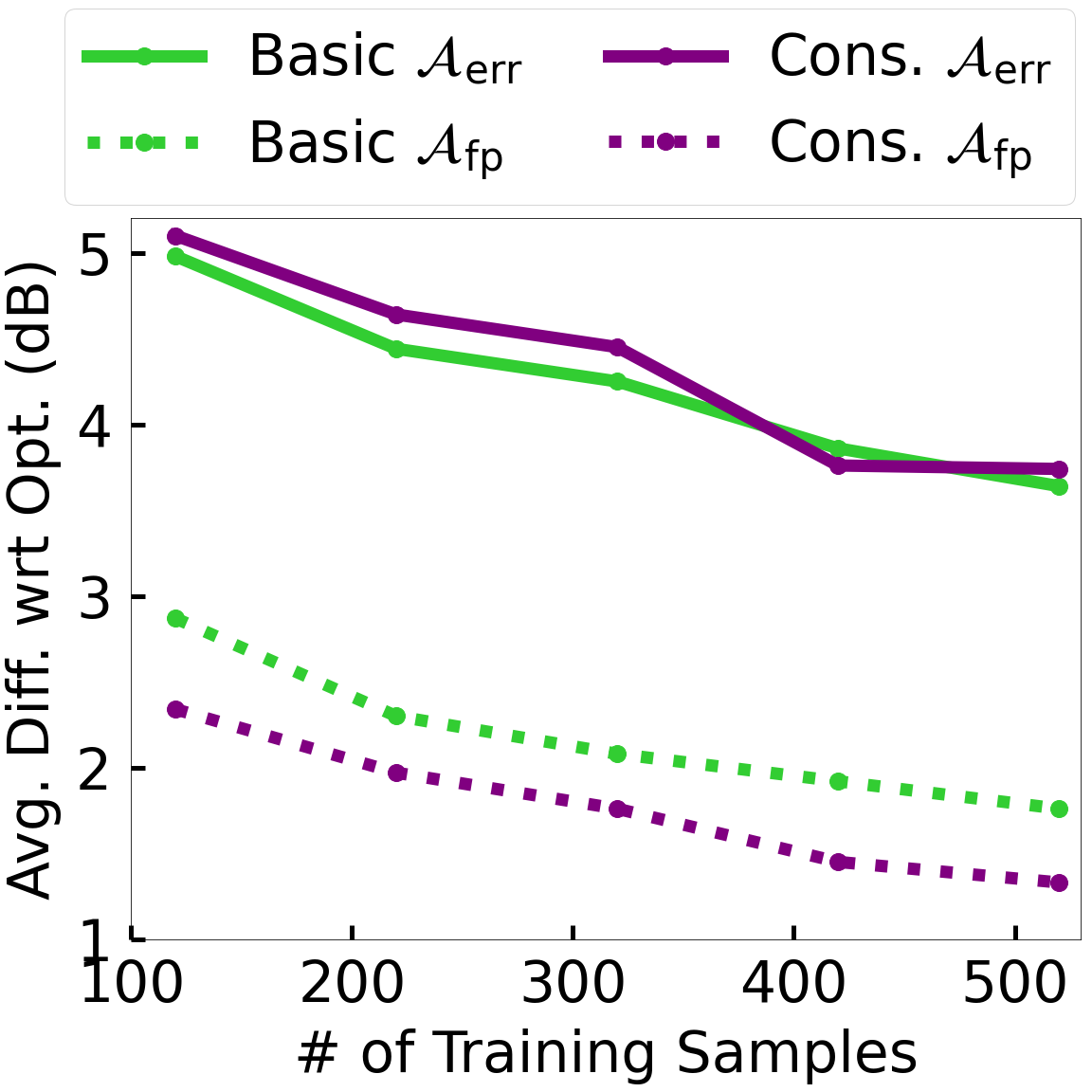} }}%
    \qquad\hspace{-0.16in}
    \subfloat[Minimizing training cost via synthetic samples in \pset.]{{\includegraphics[width=.225\textwidth]{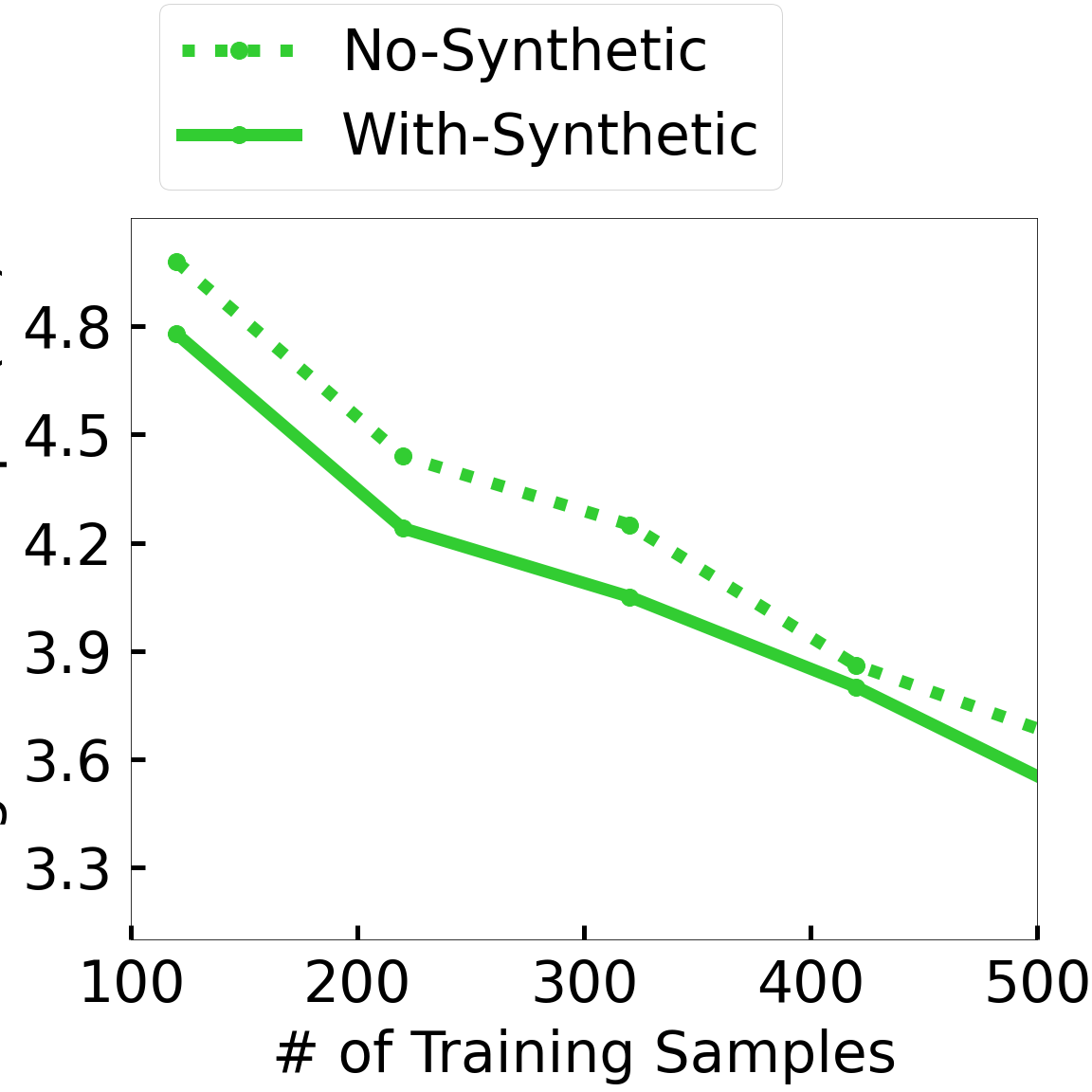} }}%
    \qquad\hspace{-0.16in}
    \subfloat[Minimizing training cost via synthetic samples in \sset.]{{\includegraphics[width=.225\textwidth]{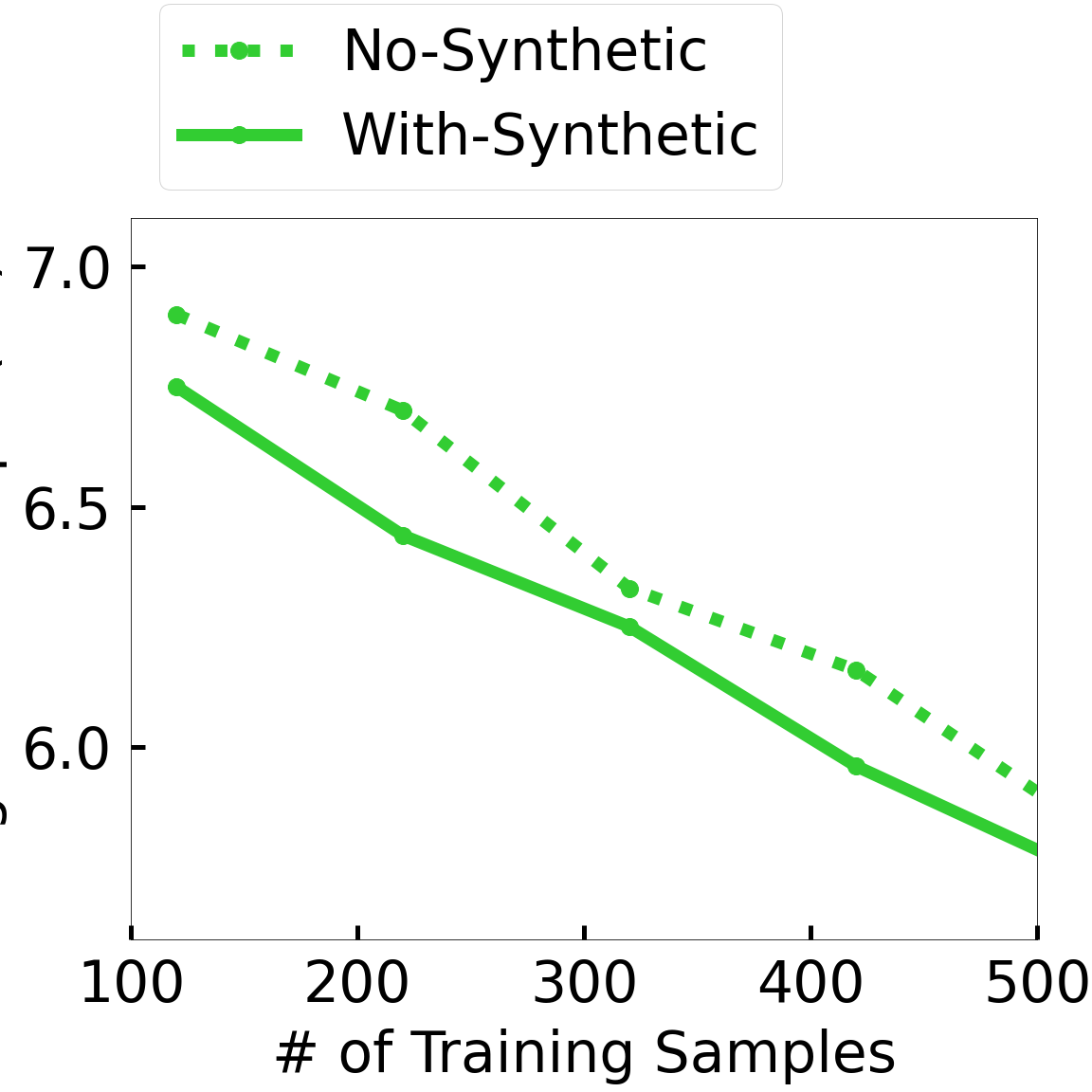} }}%
    \caption{Evaluation of various aspects in the testbed.}
    \label{fig:testbed_others}
    \end{minipage}
\end{figure*}

\section{\name TESTBED IMPLEMENTATION}
\label{sec:testbed}

In this section, we implement and evaluate a complete testbed system for our spectrum allocation system. We use the testbed to collect training samples, which are then used
to train  and evaluate the learned models. The testbed implementation demonstrates the
effectiveness of our techniques in a realistic small-scale setting. 

\para{Transmitter and Receiver Devices Used.} 
We use 4 USRP B200/B210 and 1 HackRF to play the role of the 4 PUs and 1 SU, respectively. 
A spectrum sensor (SS) is composed of an RTL-SDR dongle that connects to a dipole 
antenna and is powered by a single-board computer Odroid-C2. We deploy 17 of these 
spectrum sensors in the testbed. See Fig.~\ref{fig:testbed}(a). 
To simulate PU receivers (i.e., PURs), 
we use the same RTL-SDR dongle and antenna but power it by a laptop. Each PU is paired 
with one PUR, and they are both powered by a single laptop.
\blue{Overall, we implemented a Python repository running on Linux that transmits and receives signals and measures and collects relevant parameters in real-time at 
915 MHz ISM band at a sample rate of 1 MHz.  We built our custom communication system based on GNU Radio for data
communication between PU and PURs, used to determine labels (see below). See Fig.~\ref{fig:testbed_system}. }

\para{Collecting Training Samples.} Recall that a  sample in \pset is comprised of a sample of PUs' parameters (location and power) and the optimal power allocated to the SU. In \sset, a training sample is comprised of spectrum sensors' received power readings. The location of entities is available by using a GPS dongle connected to the laptops as described below, and the sensor's received power is computed as follows. First, we compute an FFT on the I/Q samples collected within a time window to get a power spectral density (PSD) plot. Then, we compute the area under the PSD curve over the 1 MHz channel of interest (see below), and finally, convert the computed area to an appropriate unit.

\softpara{Determining Labels (Optimal Power Allocated to SU).} We essentially do a binary search to estimate the optimal power that can be allocated to SU.  To determine whether PU to PUR transmission is incurring any harmful interference from SU, we have PU continuously streaming ASCII messages over the 1 MHz bandwidth channel centered at frequency 915.8 MHz, and check if the messages are successfully received at the PUR. This end-to-end communication system is implemented using GNU Radio.

\para{Testbed Area and Setting.}
Due to the ongoing pandemic, the testbed was conducted in the backyard of a private house. The whole area is $24m \times 24m$ large, 
\iab{which is similar in size to the testbeds considered 
in recent works~\cite{ipsn20-localize,splot} for shared spectrum systems.}
We divide the area into 100 grid cells where each represents $2.4m \times 2.4m$.
See Figure \ref{fig:testbed}(b).
We use a GPS dongle that returns the location in (latitude and longitude) and the program converts it into coordinates. We determine the location of PU, SU, and the sensors with the help of GPS dongles and manual
observation.
All the Odroids and laptops are connected to an outdoor WiFi router and communicate through ssh protocol.
For training and evaluation, the 17 sensing devices are placed on the ground and are uniformly spread out. PUs and SU are randomly placed in the area such that different regions of the backyard are "covered".
PUs' power is randomly assigned within a certain range. 

\eat{\para{Testbed as System:}
Deploying spectrum sensors at fixed locations, we move PUs (and their PUR) and SU around the field. Upon allocating random powers to PUs, PUs and SSs send their transmitting/received power values to SM. SM tries to find the highest power value that can be assigned to SU without interfering \iab{with} any PU transmission. We collect our data samples this way for both \pset and \sset at the same time. After collecting a sufficient amount of samples, SM will learn the best model for future prediction. Now, when an SU request comes, SM will check if PU information is present. If so, it uses the \pset model as it performs a higher performance. Otherwise, it uses the trained \sset model for prediction. Predicting SU highest power, SM will send a message to SU of this value. 
}

\eat{\begin{wrapfigure}{r}{1.7in}
    \vspace*{-0.2in}
    \hspace*{-0.2in}
    \includegraphics[width=1.1\linewidth, center]{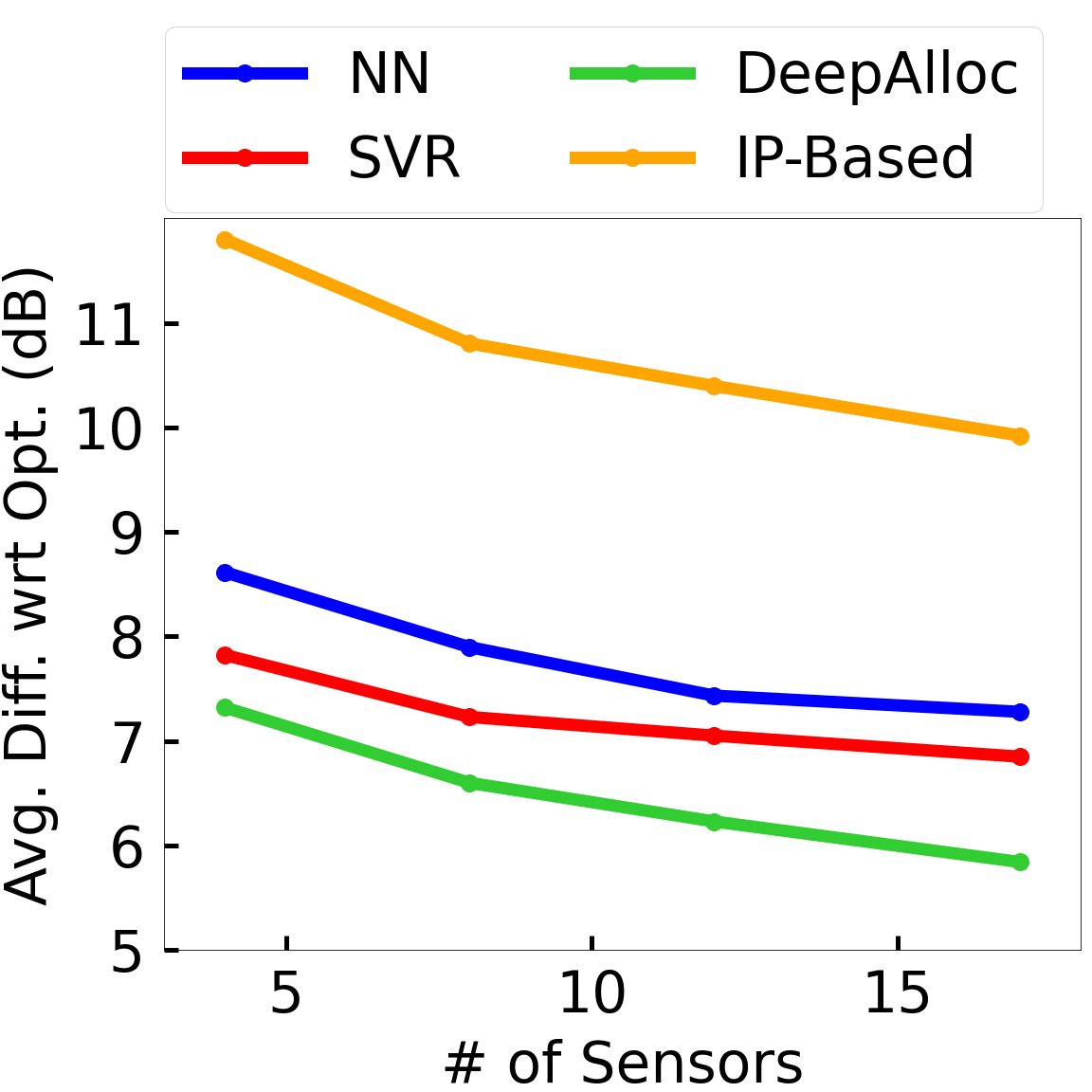}
    \vspace*{-0.1in}
    \caption{Testbed performance in \sset with increasing number of SSs and 520 training samples.} 
    \label{fig:testbed_ss_dense}
    \vspace*{-0.2in}
\end{wrapfigure}}
\para{Results.}
First, we evaluate the performance of an increasing number of training 
samples. See Fig. \ref{fig:testbed_training_size}. We observe a similar trend as in the
previous section of large-scale simulation, with \name outperforming other algorithms with
a notable margin. In particular, the overall performance of \name is good, with
4-5dB error using only 500 training samples.
The performance of all algorithms is better in \pset relative to the \sset,
as in \S\ref{sec:simulations}. Note that \ip performs quite poorly in this realistic setting,
in spite of having knowledge of PU information as well as SS readings, as it assumes an imprecise
propagation model.
%
Finally, we evaluate our techniques to handle various aspects, viz., false positive error, multi-path effect, and synthetic samples in Fig.~\ref{fig:testbed_others}. Overall, we observe a similar trend as in the large-scale simulations.

\section{Conclusions}
\label{sec:conc}

We have developed an effective deep-learning technique based on CNNs to 
learn the spectrum allocation function, and have demonstrated its effectiveness 
via extensive large-scale simulations as well as a small outdoor testbed. 
There are many avenues for further improvement of our techniques. First, one could easily
pre-train the \name model with a much larger number of pre-training samples; as our 
pre-training samples can be
automatically generated, this only incurs additional computational cost but no additional
field training/deployment cost. Second, we 
could also use more sophisticated models to generate high-fidelity pre-training samples.
Lastly, we could incorporate terrain information in the image sheets to aid the learning
process. These directions form the focus of our future work.

\bibliographystyle{plain}
\bibliography{main}



\begin{IEEEbiography}[{\includegraphics[width=1.1in,height=1.3in,clip,keepaspectratio]{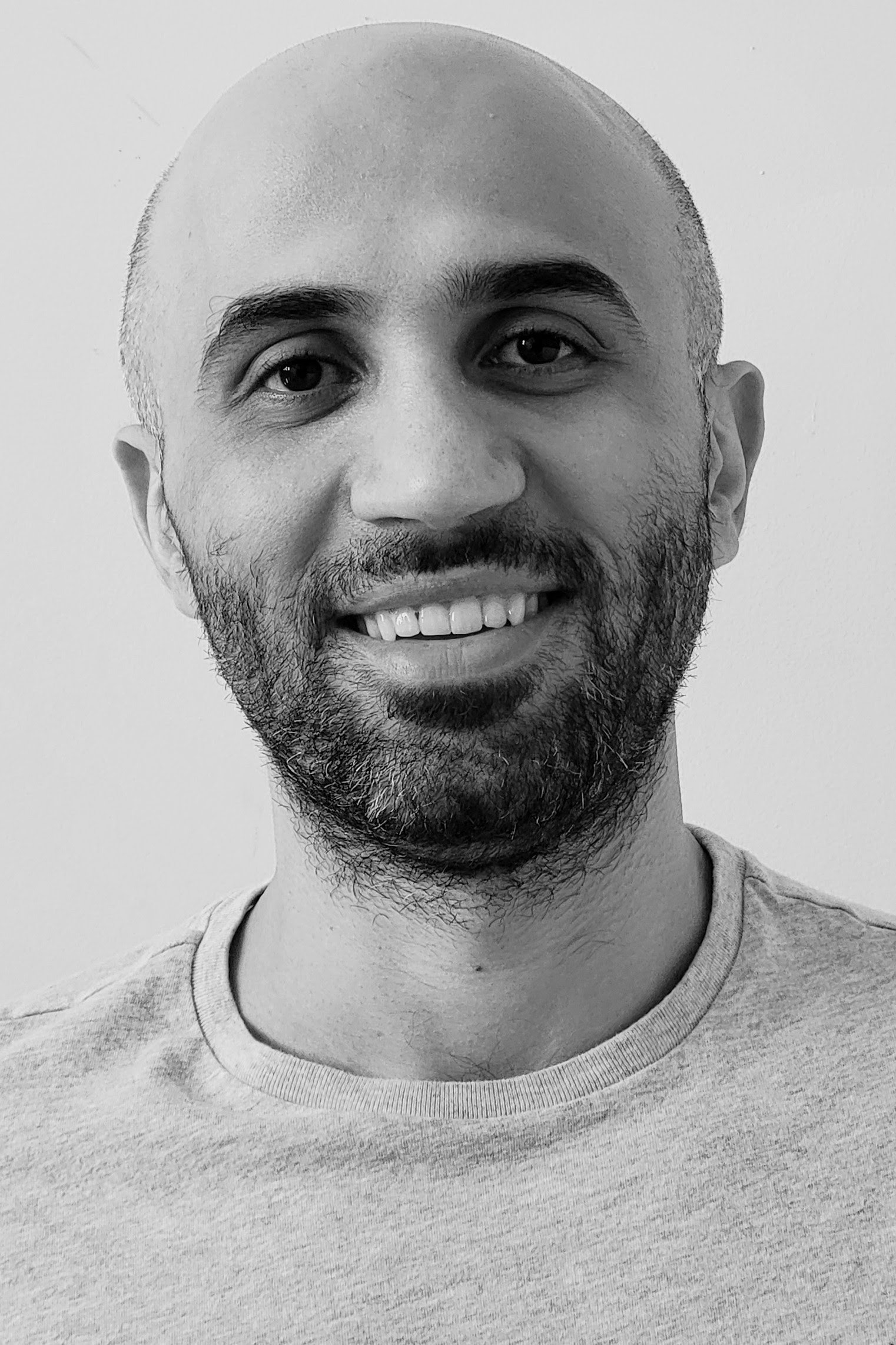}}]{Mohammad Ghaderibaneh} received the B.S. degree in electrical engineering from Azad University, Urmia, Iran in 2008 and the M.S. degree in telecommunication engineering from Shahid Beheshti University, Tehran, Iran in 2011. He received his Ph.D. degree in computer science at Stony Brook University, NY, USA in 2023.

From 2012 to 2018, he worked in the industry as a Software and Telecommunication Engineer developing wireless communication networks like TETRA. He now works at Google YouTube. His research interests include quantum networks and machine learning application in wireless networks.
\end{IEEEbiography}

\begin{IEEEbiography}[{\includegraphics[width=1.1in,height=1.3in,clip,keepaspectratio]{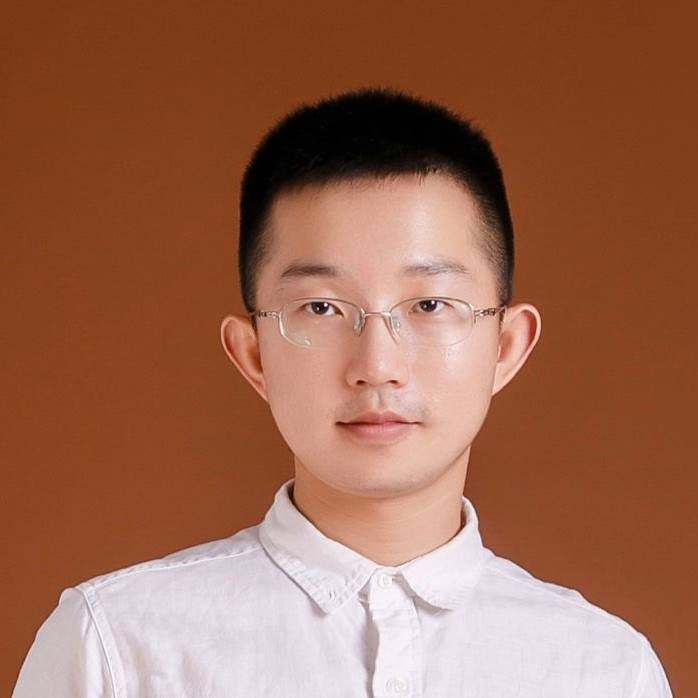}}]%
{Caitao Zhan}
received his B.S. degree in computer science and technology from China University of Geosciences in Wuhan, China in 2017. He then joined the PhD program at Stony Brook University at the Department of Computer Science and is now a PhD candidate. He does research in the broad area of computer networks and his interest lies at the intersection of wireless networks and machine learning. He is also interested in quantum communication and quantum sensing.
\end{IEEEbiography}

\begin{IEEEbiography}[{\includegraphics[width=1.1in,height=1.3in,clip,keepaspectratio]{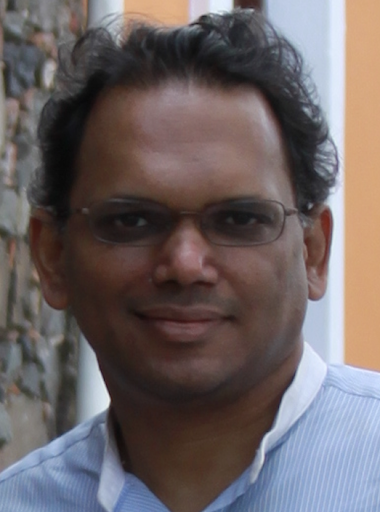}}] {Himanshu Gupta}
is a professor of computer science at Stony Brook University, where he has been a faculty since 2002. His area of research has been in wireless networks, with recent focus on free-space optical communication networks and spectrum management. His current research focuses on quantum networks and communication, and distributed quantum algorithms. He graduated with an M.S. and Ph.D. in Computer Science from Stanford University in 1999, and a B.Tech. in Computer Science and Engineering from IIT Bombay in 1992.
\end{IEEEbiography}

\end{document}